\documentclass[useAMS,usenatbib]{mn2e}
\usepackage[dvips,usenames]{color}
\usepackage{graphicx}
\usepackage{amssymb}
\usepackage{amsmath}

\newcommand{\uv}{\textbf{v}} 
\newcommand{\td}{\mathrm{d}} 

\title[Global flow and local fluctuations in 3D SPH discs]{Characterisation of global flow and local fluctuations in 3D SPH simulations of protoplanetary discs}
\author[S.~E.~Arena and J.-F.~Gonzalez]{S.~E.~Arena,$^{1}$ J.-F.~Gonzalez,$^{1}$\\
$^{1}$ Universit\'e de Lyon, Lyon,F-69003, France;
Universit\'e Lyon 1, Observatoire de Lyon, 9 avenue Charles Andr\'e, Saint-Genis Laval,\\ F-69230, France;
CNRS, UMR 5574, Centre de Recherche Astrophysique de Lyon;
\'Ecole Normale Sup\'erieure de Lyon, F-69007, France
}

\begin{document}

\date{Accepted 2013 April 22. Received 2013 April 12; in original form 2013 February 08}

\pagerange{\pageref{firstpage}--\pageref{lastpage}} \pubyear{2013}

\maketitle

\label{firstpage}

\begin{abstract}
A complete and detailed knowledge of the structure of the gaseous component in protoplanetary discs is essential to the study of dust evolution during the early phases of pre-planetesimal formation.
The aim of this paper is to determine if three-dimensional accretion discs simulated by the Smoothed Particle Hydrodynamics (SPH) method can reproduce the observational data now available and the expected turbulent nature of protoplanetary discs.
The investigation is carried out by setting up a suite of diagnostic tools specifically designed to characterise both the global flow and the fluctuations of the gaseous disc. 
The main result concerns the role of the artificial viscosity implementation in the SPH method: in addition to the already known ability of SPH artificial viscosity to mimic a physical-like viscosity under specific conditions, we show how the same artificial viscosity prescription behaves like an implicit turbulence model. In fact, we identify a threshold for the parameters in the standard artificial viscosity above which SPH disc models present a cascade in the power spectrum of velocity fluctuations, turbulent diffusion and a mass accretion rate of the same order of magnitude as measured in observations.
Furthermore, the turbulence properties observed locally in SPH disc models are accompanied by meridional circulation in the global flow of the gas, proving that the two mechanisms can coexist.

\end{abstract}


\section{Introduction}
\label{sec:intro}

Protoplanetary discs (abbreviated PPD) 
are discs composed mainly of gas and dust
in quasi-Keplerian rotation around young stars.
They are believed to be the birth place of planets.
The theory of planet formation is complex because 
several scales (from $\mu$m to hundreds of thousands of km),
several forces  (e.g. electrostatic, magnetic, gravitational, drag)
and several processes (e.g. turbulence, chemical and thermodynamical 
transformations, instabilities) are involved.
One of the more poorly understood stages is the evolution of
 $\mu$m-size dust grains into km-size objects, called planetesimals.

Among the several mechanisms the dust is subject to 
(e.g. radial drift and vertical settling), 
possible large scale motion and turbulence
are expected to be of relevant importance.
\emph{Large scale motion} can lead dust to travel 
to very different locations in the disc. 
\emph{Turbulence} can have two competing 
effects mediated by gas drag: stirring up 
and diffusing dust particles, impeding their agglomeration,
 or trapping them inside eddies, favoring their agglomeration
\citep[see e.g.][]{Cuzzi1993,Carballido2008,Cuzzi2008}. 

Here we focus on the \emph{global flow} and on the 
expected \emph{turbulent behaviour} of only 
 the \emph{gaseous} component of PPD, which represents
the medium in which dust evolves.

\defcitealias{Shakura1973}{SS73}
In the literature, two large classes of models of turbulent discs
are widely used: continuous and discrete models.
\emph{Continuous models} of viscous discs
are directly derived from the Navier-Stokes equations
in presence of the gravitational potential of the central
star, therefore they include the ingredients of 
\emph{physical viscosity} and \emph{star gravity}.
Among the  most popular models of this type 
we find the analytic 1D models 
by \cite{Pringle1981} and \cite{Lynden-Bell1974}.
Since the physical molecular viscosity of the gas
in PPD is very low 
\citep[see e.g.][]{Armitage2007}, high Reynolds numbers are
expected and  therefore the gas is believed
to be turbulent. The most commonly used model for such discs
is the    
\cite{Shakura1973} model (hereafter SS73),
which is a Prandtl model for turbulence:
 the Navier-Stokes equations are averaged 
and the Reynolds stress tensor is modeled by 
a viscosity called \emph{turbulent viscosity}.
The resulting total viscosity is the sum of 
the molecular and the turbulent viscosities, 
the former being negligible in PPD.
Continuous models are usually applied to turbulent discs 
whose source of turbulence is not known or 
not explicitly declared; in such cases the
\citetalias{Shakura1973} relation  
$\nu_{\mathrm{\scriptstyle{T}}} = \alpha_{\mathrm{\scriptstyle{ss}}}  c_\mathrm{s} H$
is used to express the turbulent viscosity parameter.
 The dimensionless parameter
 $\alpha_{\mathrm{\scriptstyle{ss}}}$, usually taken as a constant, 
collects all the ignorance of the source of turbulence,
  $c_\mathrm{s}$ is the gas sound speed and $H$ the scale height of the disc.
Generally, $\alpha_{\mathrm{\scriptstyle{ss}}}$-discs are good models for turbulence and capture its main effects. 
An example is given by the two-dimensional models calculated by
  \citet[hereafter TL02]{Takeuchi2002} that show how the $\alpha_{\mathrm{\scriptstyle{ss}}}$ prescription is able to describe the mechanism of meridional circulation in discs.
However, turbulent fluctuations are not directly reproduced in such disc
models, only their effect on the dynamics of the global discs can be studied.
In addition, if necessary, 
turbulent diffusion has to be added to the basic equations.
\defcitealias{Takeuchi2002}{TL02}

A different line of research starts from a proposed explicit 
hypothesis concerning the 
source of turbulence and derive the corresponding 
disc evolution. Given the complex dynamics, numerical simulations are
often required, therefore these consist of
mainly \emph{discrete models}. 
Turbulent fluctuations can be resolved down to the resolution scale, 
which limits the size of the smallest structures.
A first example of such models are discs produced by
Magneto-hydrodynamics (MHD) simulations where the 
source of turbulence is the Magneto-Rotational Instability (MRI)
\citep[e.g.][]{Fromang2006}.
The equations at the base of such models are the Euler equations 
in presence of an adequate magnetic field and of 
the gravitational potential of the central star.
Therefore they include the ingredients of
\emph{magnetic field} and \emph{star gravity}, 
however physical viscosity (included in continuous models)
is not present. 
In addition, \emph{turbulent diffusion} (not added a priori) 
has been found in the outcome of simulations.
A second example of discrete models are systems produced by
simulations of self-gravitating discs where the source of turbulence is
the Gravitational Instability (GI)  
 \citep[e.g][]{Rice2005}.
The equations at the base of such models are the Euler equations 
in presence of both self-gravity and 
the gravitational potential of the central star.
In addition, the numerical scheme is completed by an artificial viscosity 
term.
Therefore they include the ingredients of
\emph{self-gravity}, \emph{star gravity} and 
a \emph{`physical-like' viscosity}. 

The nature of the source of turbulence in accretion discs
and particularly in PPD is still an open issue.
Even in the absence of magnetic fields and self-gravity,
other mechanisms can be sources of turbulence, 
like interactions with external objects or 
pure  hydrodynamic and thermal instabilities, such as
the Rossby instability \citep[][]{Lovelace1999} or 
the baroclinic instability \citep[][]{Klahr2003}.
The possibility to have a hydrodynamic instability in 
accretion discs is still debated.
In fact, due to the Rayleigh criterion \citep[see e.g.][]{Balbus1991,Armitage2007},
Keplerian discs are believed to be stable with respect to
hydrodynamic instabilities.
However this is proved only for 1D discs,
if the third dimension is considered the behaviour could change and 
there may be the possibility to trigger instabilities
\citep[see e.g.][]{Nelson2012}.

As done in most continuous analytic models,
we prefer to model the effects of turbulence 
(and not its cause) by the use of an \emph{effective viscosity}.
Here we consider three-dimensional accretion discs
modeled by means of standard Smoothed Particle Hydrodynamics 
\citep[SPH, for a review see][]{Monaghan2005},
which we call \emph{SPH disc models},
they are discrete models, such as those of the second class.
The SPH scheme includes an artificial viscosity term
(a Von Neumann-like viscosity)
that has been introduced to treat shocks correctly 
\citep[see e.g.][]{Cullen2010}. 
Far from the shock, the artificial viscosity is 
usually turned off by a switch.
Different flavours of the artificial viscosity
term are discussed in Sect.~\ref{subsec:SPHcode}.

The aim of the present paper is to answer the following question:
\emph{can SPH disc models correctly reproduce both 
the observed properties of PPD and the expected effect of
turbulence?}
In order to clarify these two points we present a detailed 
characterisation of the properties of the \emph{global flow} and
particularly of the density and 
velocity \emph{fluctuations} of SPH disc models.

The global flow of SPH models gives us information about the average 
properties of the gaseous disc, while the divergence from the average
values defines the SPH fluctuations associated to the considered field
(e.g. velocity and density fields).
In particular, \emph{SPH fluctuations} result from the combination
of standard \emph{numerical noise} and 
\emph{physical fluctuations}.
The former, present in all numerical schemes, 
is due in the case of the SPH method both to the discretisation of the
continuum equations and to the SPH approximations;
it depends on the number of particles and
 on the SPH kernel used in the simulations
(see Sect.~\ref{subsec:SPHcode}).
The latter is present if the gas flow is turbulent;
it depends on the Reynolds number of the flow which in turn
is related to the physical viscosity of the gas.
Both the numerical noise and physical fluctuations
are related to the artificial viscosity term in 
the specific way we will show in this paper.

Turbulence modeling with the SPH method has recently become an active
field of research \citep[see][and references therein]{Monaghan2005}.
The first effort of implementing turbulence models in the SPH
equations \citep[e.g.][]{VioleauIssa2007}
is now moving in the direction of quantifying the 
ability of the SPH method to intrinsically reproduce
turbulence \citep[e.g.][]{Ellero2010, Monaghan2011}.
A comparison between SPH and grid methods is presented
in \cite{Price2010} and a detailed and clear relation between 
turbulence and resolution in the SPH  method is explained in
\cite{Price2012}.
In those studies, isotropic homogeneous turbulence 
 in 2D  boxes with periodic boundary conditions and 
with negligible gravity effects are considered.
However, the case of \emph{anisotropic} and 
\emph{inhomogeneous turbulence}
with \emph{free boundary condition} such as in 3D gaseous discs,
where also the action of gravity from the central star is relevant, 
has been addressed only in a few works \citep[][]{Murray1996, Lodato2010}, 
but without considering the global flow and the statistics of the velocity and
density field and the match with the available observations
for PPD. The present work aims to contribute
in this direction.

In Sect.~\ref{sec:sphmodels} we describe the main features of the 
SPH code we have used to model the accretion discs considered in the 
present work, the adopted reference disc model and the 
simulations of the different models we performed.
In Sect.~\ref{sec:diagnostic} we present the diagnostics applied to 
characterise both the global gas flow and the fluctuations
of the velocity field of the gas, along with the known 
reference values for PPD. We then show the results of the
applied diagnostics respectively for the \emph{global flow}
(Sect.~\ref{sec:globalflow}), the \emph{magnitude of SPH fluctuations}
(Sect.~\ref{sec:magnoise}) and the \emph{structure of SPH fluctuations}
(Sect.~\ref{sec:strunoise}).
The results are discussed in Sect.~\ref{sec:conclusions} where
we also draw our conclusions.

\section{SPH models of accretion discs}
\label{sec:sphmodels}
In this work we focus on a PPD described by given
analytic initial density and locally isothermal sound speed
profiles. The system is not homogeneous nor isotropic. 
We calculate the quasi-stationary state of such a disc
using different values of the three numerical parameters
$N$ (number of particles) and $\alpha$ and $\beta$
(artificial viscosity parameters) controlling the  SPH fluctuations, 
in order to characterise their effects.

In this section we present the code and the adopted units,
the reference disc model and the sets of simulations we performed.

\subsection{The SPH code}
\label{subsec:SPHcode}
We use the two-phase SPH code described in
\cite{Barri`ere-Fouchet2005}.
The two phases represent gas and dust that interact via aerodynamic drag.
The gas is described by the Euler equations and artificial viscosity.

Here we consider only the gas phase because we are interested in the 
characterisation of the global gas flow and its
velocity and density fluctuations. 
Thus,  the momentum conservation equation takes the form:
\begin{equation}\begin{array}{r@{\ }l}
\displaystyle\frac{\td \uv_a}{\td t} =& - \displaystyle\sum_b m_b \left( \frac{P_a}{\rho_a^2} + \frac{P_b}{\rho_b^2} + \Pi_{ab}  \right) \nabla_a W(r,h) \\[3ex]
& - \displaystyle\frac{GM_*}{\left( r_a +  \epsilon h  \right)^{3/2}} \textbf{r}_a,
\end{array}\end{equation}
where the usual SPH approximation of replacing integrals with sums over a finite number of particles $N$ has been performed.
The term $\textbf{v}_a$ is the velocity of SPH particle $a$, $t$ the time, 
$m$ the SPH particle mass, $P$ the pressure, $\rho$ the mass density,
$\textbf{r}_a$ the position of particle $a$ with respect to the central star
of mass $M_*$,  $\epsilon=0.1$ a parameter used to prevent singularities
and $W(r,h)$ the SPH smoothing kernel, with $r=|\textbf{r}_a-\textbf{r}_b|$ 
the distance between the particle $a$ and its neighbour $b$, and 
$h$ the smoothing length.
Here we use a cubic-spline kernel truncated at $2h$:
\begin{equation}
\label{eq:kernel}
W(q) = \frac{\sigma}{h^d} \left\{
  \begin{array}{ll}
    \displaystyle\frac{1}{4}(2-q)^3 - (1-q)^3 & 0 \leq q < 1;\\[2ex]
    \displaystyle\frac{1}{4}(2-q)^3 & 1 \leqslant q < 2;\\[2ex]
    0 & q \geqslant 2;
  \end{array},
\right.
\end{equation}
where $q=r/h$, $d$ is the dimension of the simulation 
(in this work $d$~=~$3$) and 
$\sigma$~=~$[2/3,10/7\pi,1/\pi]$ is the normalization constant 
respectively in 1, 2 and 3 dimensions.
The effect of different kernels will be addressed in a future work.

The smoothing length $h$ is variable and is derived from the density $\rho$:
\begin{equation}
h = \eta \left( \frac{m}{\rho}  \right)^{1/3}
\end{equation}
with $\eta=1.14$.
This choice of $\eta$ guarantees a roughly constant number of neighbours
(for 3D simulations: $N_\mathrm{neigh}$~$\approx$~$50$).
The smoothing length defines the resolution of the SPH simulations: for a larger number of particles $N$ the target number of neighbours is reached inside a smaller volume (of radius $2h$ from Eq.~\ref{eq:kernel}), implying a smaller smoothing length.
The gas is described by a locally isothermal equation of state.
The artificial viscosity term $\Pi_{ab}$ is described in
Sect.~\ref{subsec:artviscterm}.

\subsubsection{The artificial viscosity term}
\label{subsec:artviscterm}
\defcitealias{Monaghan1983}{MG83}
Two different implementations of the artificial viscosity term are
considered.
The first one, originally introduced by \cite{Monaghan1983} (hereafter MG83) 
and subsequently refined by \cite{Lattanzio1985,Monaghan1985,Monaghan1992},
is defined by:
\begin{equation}
\label{eq:AVMG83}
\Pi_{ij} = \left \{
\begin{array}{lll}
\displaystyle\frac{1}{\overline{\rho}_{ij}} {\left( -\alpha \overline{c}_{ij} \mu_{ij} + \beta \mu_{ij}^2 \right)}  & \mathrm{if} & \textbf{v}_{ij} \cdot \textbf{r}_{ij} < 0 \\
0 & \mathrm{if} & \textbf{v}_{ij}\cdot \textbf{r}_{ij} \geqslant 0 
\end{array}  \right.,
\end{equation}
where $r_{ij}$ and $v_{ij}$ are respectively the relative distance
and relative velocity of particles $i$ and $j$. 
The overlined quantities are averages between particle $i$ and its neighbouring particle $j$: 
 $\overline{c}_{ij}=(c_i+c_j)/2$, $\overline{\rho}_{ij}=(\rho_i + \rho_j)/2$, $\overline{h}_{ij}=(h_i+h_j)/2$.
Finally, 
\begin{equation}
\mu_{ij} = \frac{\overline{h}_{ij} \textbf{v}_{ij}\cdot \textbf{r}_{ij}}{\textbf{r}^2_{ij}+ \epsilon^2 \overline{h}_{ij}^2 },
\end{equation}
with $\epsilon^2=10^{-2}$.
Different combinations of the two artificial viscosity parameters have been used so far in different applications, in particular the combination $(\alpha,\beta)=(1,2)$ has been claimed to give good results \citep{Monaghan1992,Monaghan2005}. We will refer to $\alpha=1$ and $\beta=2$ as `standard' values for the artificial viscosity parameters.

\defcitealias{Murray1996}{Mu96}
The second implementation of artificial viscosity 
(hereafter identified as Mu96) is that
adopted by  \cite{Murray1996}
and \cite{Lodato2010}:
\begin{equation}
\Pi_{ij} = -
\frac{\alpha \overline{c}_{ij} \mu_{ij}}{\overline{\rho}_{ij}}.
\end{equation}
In contrast to the \citetalias{Monaghan1983} implementation, here the artificial
viscosity is applied to all particles 
(the switch present in the \citetalias{Monaghan1983} version is not applied)
 and the $\beta$ term
is set to zero. 
Therefore, discs with the standard \citetalias{Monaghan1983}
artificial viscosity implementation are naturally half as viscous 
as those with the \citetalias{Murray1996} implementation.

\subsubsection{The link to physical viscosity}
\label{subsec:physicavisc}
The reason of the \citetalias{Murray1996} choice is to better match the
conditions under which the artificial viscosity 
can represent a physical viscosity.
In fact,  \cite{Meglicki1993} 
 showed that  in the 3D continuum limit 
the \citetalias{Murray1996} artificial viscosity
has the form of a physical (shear and bulk) viscous force.
The shear term is equivalent to an $\alpha_{\mathrm{\scriptstyle{ss}}}$ 
viscosity 
(that we call $\alpha_{\mathrm{\scriptstyle{cont}}}$, where 
the subscript `cont' refers to continuum) given by:
\begin{equation}
\label{eq:alphaSPH}
\alpha_{\mathrm{\scriptstyle{cont,Mu96}}}(R)=\frac{1}{10} \alpha \frac{ \langle h \rangle_{\theta z}(R)}{H(R)},
\end{equation}
where the average on $h$ is taken along both 
the azimuthal and vertical directions
and the semi-thickness $H$ of the disc is 
defined in Sect.~\ref{subsec:refdisc}
(note that this expression is valid for the cubic spline kernel).
However, $\alpha_{\mathrm{\scriptstyle{cont}}}$
depends on the resolution, because of the presence 
of the smoothing length
\citep[see][]{Lodato2010}.

The SPH equations with the \citetalias{Murray1996} artificial viscosity
 and with a very large number of particles
(high resolution)
are therefore equivalent to those of a viscous fluid with 
 an $\alpha_{\mathrm{\scriptstyle{ss}}}$ 
given by   Eq.~\ref{eq:alphaSPH}.
Thus, artificial viscosity can be used to control the
effective viscosity in the disc and it is expected to be
responsible of physical fluctuations of the flow when the
gas is in the turbulent regime.

The 3D continuum limit of the \citetalias{Monaghan1983} artificial viscosity
is much more complex than the one presented
for the \citetalias{Murray1996} case and at the moment a simple relation
such as that in Eq.~\ref{eq:alphaSPH} is only available
for the $\alpha$ term 
\citep[][]{Meru2012}:
\begin{equation}
\label{eq:alphaSPHMG83}
\alpha_{\mathrm{\scriptstyle{cont,MG83}}}(R)=\frac{1}{20} \alpha \frac{ \langle h \rangle_{\theta z}(R)}{H(R)}.
\end{equation}
The factor of two between the numerical coefficients 
of Eqs.~\ref{eq:alphaSPH} and \ref{eq:alphaSPHMG83}
naturally arises from the fact that the \citetalias{Monaghan1983} 
artificial viscosity is applied only to half the
particles because of the presence of the switch (see Eq.~\ref{eq:AVMG83}).
Since the \citetalias{Murray1996} formula 
(developed much earlier than the \citetalias{Monaghan1983} formula)
is usually adopted to estimate the effective viscosity in SPH discs
independently of the particular AV implementation,
 SPH discs are generally less viscous than 
what has been considered so far.
In Sect.~\ref{subsec:effalpha} we present 
a method to determine the effective viscosity of an SPH
disc with a generic artificial viscosity implementation. 
It is tested using Eq.~\ref{eq:alphaSPH} in the \citetalias{Murray1996} case.
It can therefore be used for sampling numerically 
the $\alpha_{\mathrm{\scriptstyle{cont,MG83}}}$-$\alpha$ relation 
in the full ($\beta \neq 0$) \citetalias{Monaghan1983} case, 
through high-resolution simulations. 

We note that in the majority of SPH simulations 
the use of the artificial viscosity is still
preferred to the direct implementation of the 
viscous stress tensor to model a pure shear viscosity.
The main reason is that the latter method does not conserve
the total angular momentum exactly 
\citep[][]{Schaefer2005,Lodato2010} as the artificial viscosity term 
does. Secondarily, it is more computationally
demanding because of the presence of the second spatial derivative
of the velocities. 

\subsubsection{The units}
The internal units of the code are 
chosen to be  1 M$_\odot$ for mass,
100 au (Astronomical Units) for length and 
to give a gravitational constant $G=1$.
With these values the time unit is therefore 
 $10^3/2\pi$ yr.

\subsection{The reference disc model}
\label{subsec:refdisc}
We consider a typical T Tauri disc of mass 
$M_{\mathrm{\scriptstyle{disc}}}=0.01\ M_\star$
orbiting around a one solar mass star 
($M_\star=1$~M$_\odot$).
It extends from $R_\mathrm{in}=20$ to $R_\mathrm{out}=400$~au and is
characterised by
a density profile given, in cylindrical coordinates, by
\begin{equation}
\label{eq:rhoRz}
\rho(R,z) = \rho_0 \left(\frac{R}{R_0}\right)^{-s}
\exp\left(-\frac{z^2}{2H^2}\right),
\end{equation}
(expansion at small $z/H$, see \citealt{Laibe2012} for the rigorous expression)
with $\rho_0=\rho(R_0,0)$, and $s>0$.
\begin{equation}
H(R)=\frac{c_\mathrm{s}(R)}{\Omega(R)}
\end{equation}
is the semi-thickness of the disc,
related to the sound speed $c_\mathrm{s}$ and the angular velocity $\Omega$.
The disc is locally isothermal with a temperature radial profile given by
\begin{equation}
T(R) = T_0 \left(\frac{R}{R_0}\right)^{-q}
\end{equation}
with $q>0$, which leads to the sound speed profile
\begin{equation}
c_\mathrm{s}(R) = c_{\mathrm{s}0} \left(\frac{R}{R_0}\right)^{-q/2}.
\end{equation}
Note that the sound speed coefficient $c_{\mathrm{s}0}$ and 
the sound speed exponent $q/2$ determine respectively 
the semi-thickness of the disc and its radial dependence: 
\begin{equation}
H(R)=H_0 \left(\frac{R}{R_0}\right)^{\frac{3-q}{2}},
\end{equation}
with $H_0=c_{\mathrm{s}0} R_0^{3/2}/\sqrt{GM_\star} $.
The resulting radial profile of the surface density is
\begin{equation}
\Sigma(R) = \Sigma_0 \left(\frac{R}{R_0}\right)^{-p},
\end{equation}
with $\Sigma_0=\sqrt{2 \pi} \rho_0 H_0$ the surface density at $R_0$ and 
\mbox{$p=s+(q-3)/2$}.

When only the gas phase is considered, all models are self-similar
and different physical scales correspond to the same
dimensionless model.
Therefore, in the following, results are mainly expressed in code units.

The reference values we adopt in the following are 
$(p,q)=(3/2,3/4)$ and $R_0=100$~au.
At this location, the disc is slightly flared with $H_0/R_0=0.05$,
$\Sigma_0 \approx 4.58$~kg\,m$^{-2}$ and
$c_{\mathrm{s}0} \approx 149$~m\,s$^{-1}$ ($T_0 \approx 6$~K).
For reference, the corresponding value at 1~au are: 
$\Sigma_0 \approx 4580$~kg\,m$^{-2}$ and $T_0 \approx 198$~K.

We follow the evolution of the reference disc model,
after pressure equilibrium  has been reached,
up to time $t=100$ in code units (corresponding to 15.9 orbits at 100 au).
All figures are plotted for that time, unless stated otherwise.

\subsection{The simulations}
The simulations we have performed are listed in
Table~\ref{table1}.
The first column gives the name of the simulation,
the second and third  columns display the values of the two artificial
viscosity (AV) parameters $\alpha$ and $\beta$, 
the fourth the number $N$ of SPH particles used for sampling the disc,
the fifth the kind of artificial viscosity.
Simulations can be divided in five sets.
Each simulation can belong to more than one set.
Sets are displayed in the last column of the table.
Simulations in Set $A$ and Set $B$
allow to study the effect of changing the artificial 
viscosity parameters for the \citetalias{Monaghan1983} and for the 
\citetalias{Murray1996} artificial viscosity respectively.
With simulations in Set $C$ it is possible to
study the effect of the different artificial
viscosity implementations.
In Set $D$ are collected simulations designed for 
studying the effect of resolution.
Finally, simulations in Set $E$ are used for testing the  
$\alpha_{\mathrm{\scriptstyle{cont}}}$-$\alpha$ relations presented
in Eqs.~\ref{eq:alphaSPH}~and~\ref{eq:alphaSPHMG83}.
The role of each set is summarised in Table~\ref{table2}.

\begin{table}
\renewcommand{\arraystretch}{1.3}
\caption{Simulation list} 
\label{table1}
\begin{center}
\begin{tabular}[]{l l l l l l}
\hline
Name & $\alpha$ & $\beta$ & $N$  & AV & Set\\
\hline
S1  & 0.1 & 0.0   & $2 \cdot 10^5$  & MG83 & $A$,$C$\\
S2  & 0.1 & 0.2   & $2 \cdot 10^5$ & MG83 & $A$\\
S3  & 0.1 & 0.5  & $2 \cdot 10^5$  &MG83 &$A$\\
S4  & 0.1 & 2.0   & $2 \cdot 10^5$  &MG83 & $A$\\
S5  & 0.1 & 10.0   & $2 \cdot 10^5$  &MG83 &$A$\\
\hline
S6  & 1.0 & 0.0   & $2 \cdot 10^5$  &MG83 & $A$,$C$\\
S7  & 1.0 & 2.0   & $2 \cdot 10^5$ & MG83 & $A$,$D$\\
S8  & 1.0 & 5.0   & $2 \cdot 10^5$  & MG83 &$A$,$D$\\
S9  & 1.0 & 10.0   & $2 \cdot 10^5$  & MG83 &$A$\\
\hline
S10  & 2.0 & 0.0   & $2 \cdot 10^5$ &  MG83& $A,E$\\
S11  & 2.0 & 4.0   & $2 \cdot 10^5$ &  MG83& $A$\\
S12  & 2.0 & 10.0   & $2 \cdot 10^5$ &  MG83& $A$\\
\hline
S13  & 5.0 & 0.0   & $2 \cdot 10^5$  & MG83& $A$\\
S14  & 5.0 & 2.0   & $2 \cdot 10^5$  & MG83& $A$\\
S15  & 5.0 & 10.0   & $2 \cdot 10^5$  & MG83& $A$\\
\hline
S16  & 0.1 & 0.0   & $2 \cdot 10^5$  & Mu96& $B$,$C$\\
S17  & 1.0 & 0.0   & $2 \cdot 10^5$  & Mu96& $B$,$C$,$D$\\
S18  & 2.0 & 0.0   & $2 \cdot 10^5$  & Mu96& $B$\\
\hline
S19  & 1.0 & 2.0   & $5 \cdot 10^4$  & MG83& $D$\\
S20  & 1.0 & 2.0   & $1 \cdot 10^6$  & MG83& $D$,$C$ \\
\hline
S21  & 1.0 & 5.0   & $5 \cdot 10^4$  & MG83& $D$\\
S22  & 1.0 & 5.0   & $1 \cdot 10^5$  & MG83& $D$\\
S23  & 1.0 & 5.0   & $5 \cdot 10^5$  & MG83& $D$\\
S24  & 1.0 & 5.0   & $1 \cdot 10^6$  & MG83& $D$\\
\hline
S25  & 1.0 & 0.0   & $1 \cdot 10^6$ &  Mu96& $D$,$C$,$E$\\
S26  & 5.0 & 0.0   & $1 \cdot 10^6$ &  Mu96& $E$\\
S27  & 5.0 & 0.0   & $2 \cdot 10^5$ &  Mu96& $E$\\
\hline
\end{tabular}
\end{center}
\end{table}

\begin{table}
\renewcommand{\arraystretch}{1.3}
\caption{Simulation sets} 
\label{table2}
\begin{center}
\begin{tabular}[]{l l}
\hline
Set & Description \\
\hline
$A$ &  Changing $\alpha$ and $\beta$ in \citetalias{Monaghan1983} artificial viscosity\\
$B$ &  Changing $\alpha$ in \citetalias{Murray1996} artificial viscosity\\
$C$ &  Changing the artificial viscosity model (\citetalias{Monaghan1983}, \citetalias{Murray1996})\\
$D$ &  Changing $N$\\
$E$ &  Testing the $\alpha_{\mathrm{\scriptstyle{cont}}}$-$\alpha$ relations (Eqs.~\ref{eq:alphaSPH}~and~\ref{eq:alphaSPHMG83})\\
\hline
\end{tabular}
\end{center}
\end{table}

\begin{figure}
\centering
\includegraphics[width=0.4\textwidth]{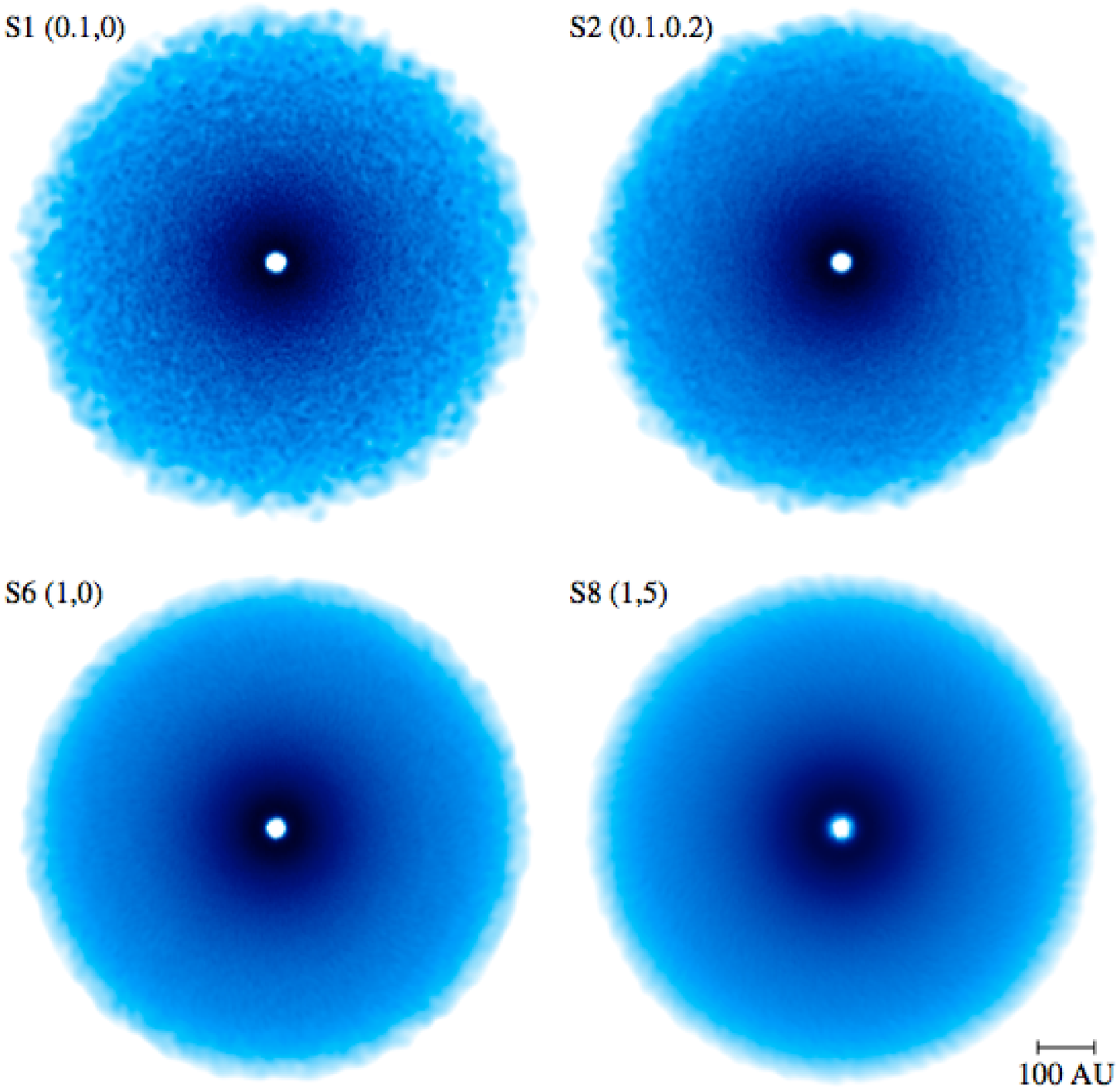}\\
\includegraphics[width=0.4\textwidth]{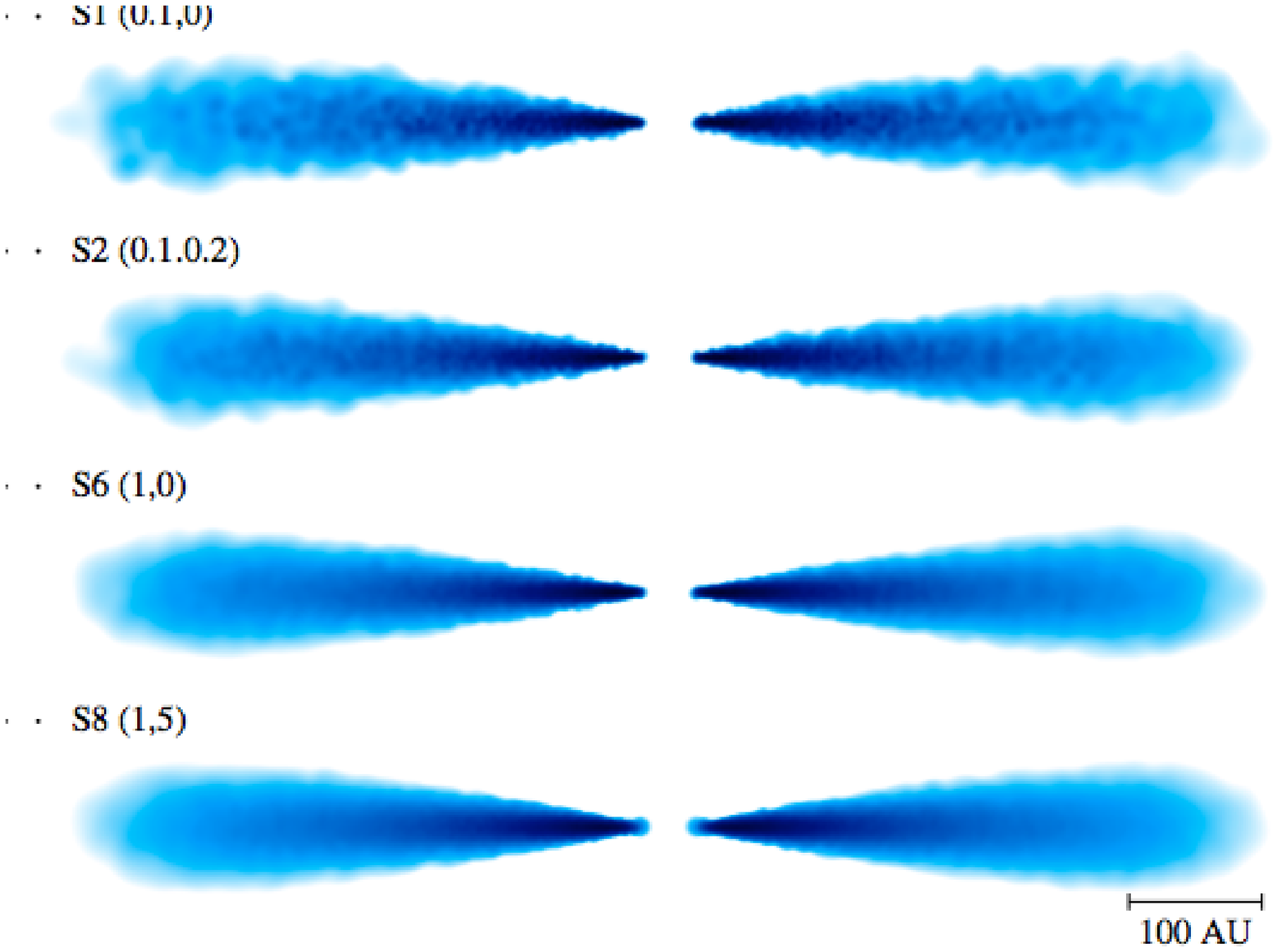}\\
\caption{\emph{Disc morphology}: volume gas density in
midplane and vertical cross sections
for simulations S1, S2, S6, S8 
(from top to bottom and from left to right).
The values of the two AV parameters 
($\alpha, \beta$) are displayed for each case.
 }
\label{fig1}
\end{figure}

Midplane and vertical cross sections of the density profiles of four
of the performed simulations are shown in Fig.~\ref{fig1}.

\section{Diagnostics and reference values for PPD}
\label{sec:diagnostic}
We consider two kinds of diagnostics which refer to the global flow 
and to fluctuations of several quantities, with particular focus on
density and velocity field.
We first define the diagnostic and then give the expected values
in some known cases.
In the following, 
the symbol $\langle \cdot \rangle$ represents  average quantities.
In this paper we perform both the standard azimuthal  and
vertical averages $\langle \cdot \rangle_{\theta z}$ 
and the azimuthal averages only $\langle \cdot \rangle_{\theta}$, 
in order to characterise also the vertical extension.

\subsection{Diagnostic for the global disc flow}
Here, we present the selected quantities used to
 characterise the global flow.

\begin{enumerate}
\item \emph{Mass distribution}. We look at:
  \begin{enumerate}
    \item the radial profile of the surface density $\Sigma(R)$,
    \item the density distribution $\rho(R_\mathrm{s},z)$ 
      in the direction perpendicular to the 
      disc midplane, for a selected radius $R_\mathrm{s}$.
  \end{enumerate}

\item \emph{Velocity structure}.  We focus on:
  \begin{enumerate}
  \item the radial velocity maps $v_R(R,z)$, where the radial component
of the velocity field is azimuthally averaged.
  \item the local Mach number of the flow in the midplane Ma$(R) \equiv v(R) / c_\mathrm{s}(R)$, where $v$ is the modulus of the local velocity field.
  \end{enumerate}

\item \emph{Macroscopic turbulence signatures}.
  Four items are analysed.
  \begin{enumerate}
  \item The mass accretion rate onto the central star $\dot{M}_0$: 
    we are particularly interested in this quantity for two reasons: 
    (1) it gives indirect information about the turbulent viscosity coefficient  and     (2) it is constrained by observations.
    For PPD around T Tauri stars the measured values of mass accretion rates  onto the central star are $\dot{M} \approx 10^{-8} $ M$_{\odot}$  \citep{Hartmann1998,Andrews2009}, they can be reproduced by
a value $\alpha_{\mathrm{\scriptstyle{ss}}}\approx 10^{-2}$
\citep{Hartmann1998,King2007}.

  \item Since $\dot{M}_0$ gives an information restricted to the 
    very inner region of the disc (where our inner boundary condition
    is free), we also analyse the radial profile 
    of the local mass accretion rate:
    $\dot{M}(R)=-2\pi R \Sigma \langle v_R \rangle_{\theta z}$,
    averaged in the azimuthal and vertical direction.

    \item The \emph{effective viscosity} $\alpha_{\mathrm{\scriptstyle{eff}}}$ of the gaseous disc (we express the viscosity in terms of the dimensionless parameter
$\alpha_{\mathrm{\scriptstyle{ss}}}$ of the \citetalias{Shakura1973} parametrization) that characterises the simulated disc.
  
The effective viscosity in the simulated disc model
is estimated by means of fits to the two-dimensional analytic models
of accretion discs 
(see details in Appendix~\ref{app:fit}).
We use 2D analytic models because they are well suited
for the azimuthal symmetry present in our simulated
discs. For this reason we call 
$\alpha_{\mathrm{\scriptstyle{2D}}}$ this way of estimating the
effective viscosity $\alpha_{\mathrm{\scriptstyle{eff}}}$.
In particular, we derive $\alpha_{\mathrm{\scriptstyle{2D}}}$ by comparing
the vertical profile of the radial velocity of the gas flow
to the expression derived by analytic 
$\alpha_{\mathrm{\scriptstyle{ss}}}$-disc models 
for different radial positions.
We rewrite the vertical profile of the radial velocity,
given by  \citetalias{Takeuchi2002} and \citealt{Fromang2011},
highlighting the scale height of the disc $H(R)$ at radial position $R$:

\begin{align}
\label{eq:vr}
\frac{v_R}{c_\mathrm{s_0}} = & 
\ \frac{\alpha_{\mathrm{\scriptstyle{2D}}} }{2} \frac{H(R)}{R_0} \left( \frac{R}{R_0} \right)^{-\frac{1+q}{2}} \times
\\[1ex] \nonumber
&
\ \left\{ 6p + q - 3 - (9-5q) \left[ \frac{z}{H(R)}  \right]^2  \right\}.
\end{align}

\cite{Lodato2010} estimated $\alpha_{\mathrm{\scriptstyle{eff}}}$ by fit
of the surface density profile of 1D viscous accretion discs
\citep{Pringle1981} and
 found good agreement with $\alpha_{\mathrm{\scriptstyle{cont,Mu96}}} $ 
in SPH simulations of
warped discs 
when the necessary equivalence conditions
(large number of particles and \citetalias{Murray1996} artificial viscosity,
see Sect.~\ref{subsec:physicavisc}) are met.

\item Once we know the effective viscosity of the disc, 
  we can derive the corresponding Reynolds number:
\begin{equation}
\mathrm{Re}_{\mathrm{\scriptstyle{eff}}}=\frac{\mathrm{Ma}}{\alpha_{\mathrm{\scriptstyle{eff}}}}.
\end{equation}
\end{enumerate}
 \end{enumerate}

\subsection{Diagnostic for fluctuations}
\label{subsec:gtf}
For each simulated disc we have studied both the magnitude and the structure of the fluctuations present in the disc with particular attention to
the velocity field.
Since current observations have not yet reached the resolution 
necessary to directly detect turbulence and study its features
in protoplanetary discs (some progress has been recently claimed by 
\citealt{Hughes2011} and \citealt{Guilloteau2012}, who measured
some turbulent velocities in mm observations)
the resulting properties of density and
velocity fluctuations have been compared to the 
typical behaviour observed in turbulence experiments 
and to results from grid- and particle-based simulations 
available in the literature.

The components of the velocity field
are identified by $v_i$ and those of the fluctuating velocity field 
are $u_i=v_i- \langle v_i \rangle$ with $i=R, \theta, z$. 

\subsubsection{Magnitude of velocity fluctuations}
Concerning the \emph{magnitude} of velocity fluctuations, 
we focus on the turbulent viscosity coefficient (point M1), 
the diffusion coefficients (point M2) and the  Mach number
of velocity fluctuations (point M3).

\begin{enumerate}
\item[(M1)] The \emph{turbulent viscosity coefficient} $\nu_{\mathrm{\scriptstyle{T}}}$
is often  derived from Reynolds averages of Navier-Stokes equations with the
 turbulent viscosity hypothesis \citep[e.g.][]{Pope2000} and
is thus related to Reynolds stresses:  
\begin{equation}
\label{nuT}
\nu_{\mathrm{\scriptstyle{T}}}=- \langle u_R u_{\theta} \rangle_{\theta} \left/ 
\left[ R \frac{\partial \left( \langle v_{\theta} \rangle_{\theta} / R \right)}{\partial R} \right]\right.
\end{equation}
(see Appendix~\ref{app:turbulentVisc}).

Here we call $\alpha_{\mathrm{\scriptstyle{RS}}}$
the corresponding \citetalias{Shakura1973} parameter for 
a disc in quasi-Keplerian rotation, it is given by:
\begin{equation}
\alpha_{\mathrm{\scriptstyle{RS}}}(R,z) =  \frac{2}{3} \frac{\langle u_R u_{\theta} \rangle_{\theta}}{c_\mathrm{s}^2},
\end{equation}
where Eq.~\ref{nuT} has been combined with the 
\citetalias{Shakura1973} relation 
$\nu_{\mathrm{\scriptstyle{T}}}$~=~$\alpha_{\mathrm{\scriptstyle{ss}}}  c_\mathrm{s} H$
(see Appendix~\ref{app:alphaRS}).
For each radial and vertical position, 
averages are performed spatially along the azimuthal
direction in order to be able to derive the radial and vertical
dependence of $\alpha_{\mathrm{\scriptstyle{RS}}}$.

The corresponding \emph{turbulent} Reynolds number is:
\begin{equation} 
\mathrm{Re}_{\mathrm{\scriptstyle{T}}}= \frac{\mathrm{Ma}}{\alpha_{\mathrm{\scriptstyle{RS}}}}.
\end{equation}

In MHD simulations of accretion discs,
$\alpha_{\mathrm{\scriptstyle{ss}}}\approx 10^{-3}$ and includes both
the Reynolds and Maxwell stresses \citep[e.g.][]{Fromang2009,Flock2011a}, 
in simulations of self-gravitating discs,
$\alpha_{\mathrm{\scriptstyle{ss}}}\approx 10^{-2}$ and includes both
the Reynolds and gravitational stresses
 \citep[e.g.][]{Rice2005,Rice2011}.

\item[(M2)] Turbulent flows are often approximated
by means of models where turbulence is
 described as a diffusive process.
Here we want to determine if diffusion is present in our simulations.
To this end the method used by \cite{Fromang2006a}
in order to calculate the \emph{turbulent diffusion coefficient}
$D_{\mathrm{\scriptstyle{T}}}$
is well suited for Lagrangian codes such as SPH:
\begin{equation}
\label{eq:diffcoeff}
D_{\mathrm{\scriptstyle{T}}}(t)=\int_0^{t} S_{zz}(t') \mathrm{d} t'
\end{equation}
where $S_{zz}$ is the vertical velocity correlation function:
\begin{equation}
\label{eq:vz}
S_{zz}(t)=\langle v_z(z,t) v_z(z_0,0) \rangle,
\end{equation}
with $z_0$ representing the vertical position
of a given particle at time $t=0$ and $z$ the vertical
position of the same particle at time $t$;
in a similar way $v_z(z_0,0)$ is the velocity of the same
particle at time $t=0$ and $v_z(z,t)$ at time $t$.
In order to compute the diffusion coefficient
at a selected radial location $R_\mathrm{s}$,
we average the diffusion coefficient of single particles
that were at $R_\mathrm{s}$ 
at the beginning of the simulation.

A diffusion coefficient
$D_{\mathrm{\scriptstyle{T}}}/(c_\mathrm{s}H) \approx 5 \cdot 10^{-3}$
is found by \cite{Fromang2006a}
in MHD local shearing box stratified simulations.

\item[(M3)] The strength of velocity fluctuations and therefore
their subsonic or supersonic behaviour can be determined by 
the corresponding \emph{Mach number}, defined as the ratio 
of the modulus of velocity fluctuations to the gas sound speed:
\begin{equation}
\mathrm{Ma}_\mathrm{f} \equiv \frac{|u|}{c_\mathrm{s}}.
\end{equation}
We have computed azimuthal-only and azimuthal and vertical averages
in order to present $R$--$z$ maps and radial profiles.

In the global MHD simulations of stratified discs of
 \cite{Fromang2006} and  \cite{Flock2011}, the
Mach number of velocity fluctuations is $\mathrm{Ma}_\mathrm{f} \approx 0.1$ 
around the midplane
and $\mathrm{Ma}_\mathrm{f}\approx 0.4$ in the surface layers.

\end{enumerate}

\subsubsection{Structure of velocity fluctuations}
\label{sec:struct}
Concerning the \emph{structure} of velocity fluctuations, 
we focus on the anisotropy vector (point S1), 
on the power spectrum (point S2) and
on the possible presence of intermittency (point S3).

\begin{enumerate}
\item[(S1)] In order to determine the isotropic or anisotropic
character of velocity fluctuations, we define the
\emph{anisotropy vector} 
$\vec{\delta}$~=~$(\delta_R, \delta_{\theta}, \delta_z)$ 
in terms of the velocity dispersion 
in the three spatial directions, in analogy with galactic dynamics
 \citep[see e.g.][]{Bertin2000}:
\begin{equation}
\delta_i \equiv 2 - \frac{\sigma_j^2 + \sigma_k^2}{\sigma_i^2},
\end{equation}
where the square of the velocity dispersion 
$\sigma_i^2$~=~$\langle ( v_i - \langle v_i \rangle)^2  \rangle$~=~$\langle u_i^2 \rangle$~=~$\langle v_i^2 \rangle - \langle v_i \rangle^2$ and 
$i$, $j$ and $k$ refer to the three cylindrical components.
Systems with anisotropy in the $i$ direction are characterised by 
a velocity dispersion of the  $i$ component that is larger 
than that of the other
two components: $\sigma_i \gg \sigma_j$ and $ \sigma_i \gg \sigma_k$.
Therefore, the $i$ component of the anisotropy vector will be
$\delta_i  \approx 2$.
Averages are performed along both the vertical and the azimuthal direction.

From the plots of \cite{Fromang2006} and  \cite{Flock2011}
we can deduce that $\sigma_R \gg \sigma_{\theta} \approx \sigma_{z}$,
therefore MHD discs are radially anisotropic and the corresponding
anisotropy vector is $\vec{\delta} \approx (2,0,0)$.
In contrast, in experiments of turbulence produced by round jets a corresponding
azimuthal anisotropy $\vec{\delta} \approx (0,2,0)$ has been observed
 \citep[e.g.][]{Pope2000}.

\item[(S2)] The \emph{power spectrum} is computed along a ring 
of selected  radius $R_\mathrm{s}$,
centred on the star and located at the selected 
vertical position $z_s$ (see Appendix~\ref{app:PDFsEtAl}):
\begin{equation}
P_i(R_\mathrm{s},z_s;k)=|\hat{u}_i(k)|^2,
\end{equation}
with $\hat{u}_i(k)$ the Fourier transform of the component $i$ of 
the velocity fluctuation vector
and $k=1/\lambda$ the wavenumber corresponding to length scale $\lambda$.
We analyse the power spectrum of the velocity field 
in order to determine if an energy cascade,
 characteristic of turbulent systems,
is present and if differences related to the position inside the
disc are relevant.

\item[(S3)] Highly turbulent flows are characterised by \emph{intermittency}
\citep[e.g.][]{Frisch1996}. This feature
implies non-Gaussian probability distribution functions (PDFs) with
corresponding non-Gaussian higher order moments.
Here we focus on the density and acceleration PDFs 
and on their $3^{rd}$ (\emph{skewness}, noted $S$)
and $4^{th}$ (\emph{kurtosis}, noted $K$) order moments
(see Appendix~\ref{app:PDFsEtAl}).
\end{enumerate}

Data are not available for the  power spectrum
of fluctuations and for the phenomenon
of intermittency in accretion discs.
In the classical Kolmogorov theory of incompressible homogeneous turbulence
the power spectrum of velocity fluctuations
has a power law form $P(k) \sim k^s$ with $s$~=~$-5/3$.
Recent simulations of supersonic compressible gas found a slope
of $s$~$\approx$~$-2$  \citep[][and references therein]{Price2010}.
Intermittency, highlighted by non-Gaussian PDFs and higher order moments,
is experimentally found in very high Reynolds number
incompressible flows.
In SPH simulations of a simple weakly compressible
periodic shear flow, \cite{Ellero2010} 
find that the PDF of the
particle acceleration is in good agreement with non-Gaussian statistics
observed experimentally for incompressible flows.

In simulations of supersonic compressible turbulence 
 \citep[see e.g.][]{Price2010}, log-normal distributions of the 
density PDF have been observed:
\begin{equation}
p(x) = \frac{1}{\sqrt{2 \pi \sigma_p^2}} \exp\left[ -\frac{(x-\langle x \rangle)^2}{2 \sigma_p^2}  \right] 
\end{equation}
with $x=\textrm{ln}(\rho/ \langle \rho \rangle)$ and 
a width controlled by the Mach number of the fluctuations
\begin{equation}
\label{eq:sigmap}
\sigma_p^2=\textrm{ln}\left( 1 + b^2 \textrm{Ma}^2 \right),
\end{equation}
where $b \approx 0.5$.
The expressions of the skewness and kurtosis
in terms of $\sigma_p$ for the log-normal distribution are
\begin{equation}
\label{eq:SK}
\left\{
\begin{array}{r c l}
S & = & \left( \textrm{e}^{\sigma_p^2}+2 \right) \sqrt{\textrm{e}^{\sigma_p^2} - 1} \\
K & = & \textrm{e}^{4\sigma_p^2} + 2 \textrm{e}^{3\sigma_p^2} + 3\textrm{e}^{2\sigma_p^2} - 3.
\end{array} \right.
\end{equation}
When $\sigma_p \rightarrow 0$ the log-normal distribution tends to
a Gaussian distribution and the skewness and kurtosis tend to their gaussian
values (0 and 3 respectively).

The results  of the application of these diagnostics
 to the simulations we performed are described in Sects~\ref{sec:globalflow},
\ref{sec:magnoise} and \ref{sec:strunoise}.

\section{The disc global flow}
\label{sec:globalflow}
We now consider how the main properties of the global disc flow
(mass distribution, velocity structure, mass accretion rate and
effective disc viscosity) depend
on the strength of artificial viscosity.
The convergence of each quantity is also studied.

\subsection{Mass distribution}
\begin{figure}
\centering
\includegraphics[width=0.23\textwidth]{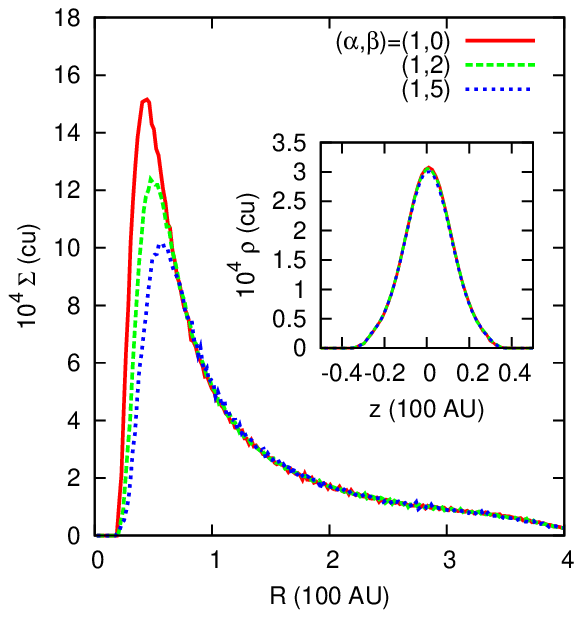}
\includegraphics[width=0.23\textwidth]{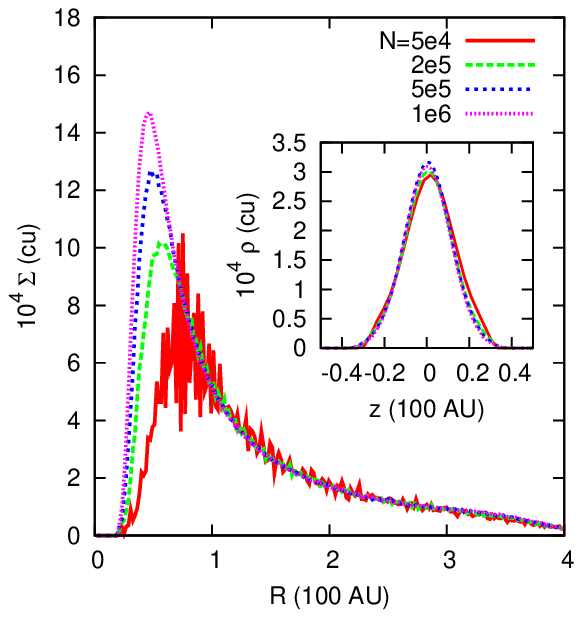}
\caption{\emph{Mass distribution.} Dependence of the surface density 
$\Sigma(R)$ and vertical density profile $\rho(R_\mathrm{s}=1,z)$
on the AV parameters 
(left: simulations S6, S7, S8, S9 with $N=2 \cdot 10^5$ particles) and 
on the resolution 
(right: simulations S21, S8, S23, S24 with ($\alpha$,$\beta$)=(1,5)). 
Values are given in code units.}
\label{figNew-01}
\end{figure}

The position of the peak 
of the surface density profile $\Sigma(R)$ moves  
towards larger radial distances and its maximum decreases 
(see left panel of Fig.~\ref{figNew-01}),  
when the two AV parameters 
$\alpha$ and $\beta$ are increased:
such behaviour is in agreement with 
the viscous spreading mechanism that is expected here
due to the existing correlation 
between artificial and physical viscosity 
(see Sect.~\ref{subsec:physicavisc}).

An increase in resolution (see right panel of Fig.~\ref{figNew-01})
has a qualitatively
similar effect to a decrease in viscosity: the surface density peak
moves towards the centre of the disc, becoming higher
(note also that the curve becomes less noisy).
In the central and external regions the projected density profile converges
already for a number of particles as small as $5 \cdot 10^4$.
In the inner region ($R < 10^{-1} R_{out}$),  
a trend towards convergence is present,
but a higher number of particles, $N \gtrsim 10^6$, is necessary.

The vertical density profile $\rho(R_\mathrm{s},z)$ at a selected radius
$R_\mathrm{s}$ (see insets in Fig.~\ref{figNew-01} for $R_\mathrm{s}=1$)
follows the expected Gaussian distribution of 
Eq.~\ref{eq:rhoRz}, showing that the desired
profile is well reproduced by the sampling procedure
and by numerical thermalisation.
It is not significantly affected by changing $\alpha$, $\beta$ or $N$.

\subsection{Velocity structure}
\label{subsec:velstruc}
\subsubsection{Radial velocity maps}
\begin{figure}
\centering
\includegraphics[width=0.5\textwidth]{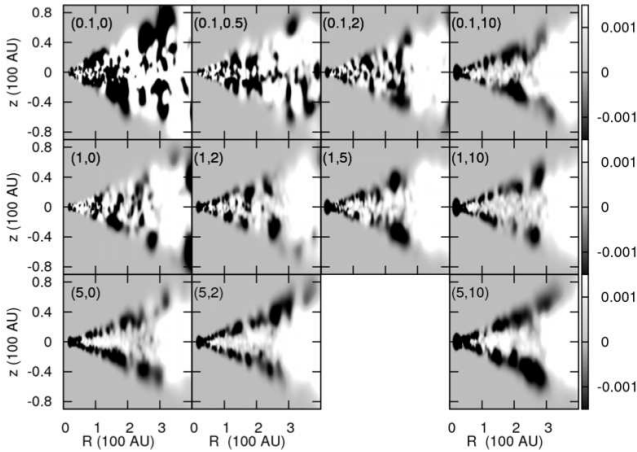}
\caption{\emph{Radial velocity structure of the gaseous disc: 
changing AV parameters}. 
Maps of the azimuthally averaged radial velocity
$\langle v_R(R,z) \rangle_{\theta}$ in the meridian plane
for simulations with $N=2 \cdot 10^5$ particles and the
\citetalias{Monaghan1983} artificial viscosity implementation.
The colorbar gives the radial velocity: negative values correspond 
to gas inflow and positive values to outflow.
Values of ($\alpha$,$\beta$) are given in each subplot.
}
\label{figNew-02}
\end{figure}

\begin{figure}
\centering
\includegraphics[width=0.5\textwidth]{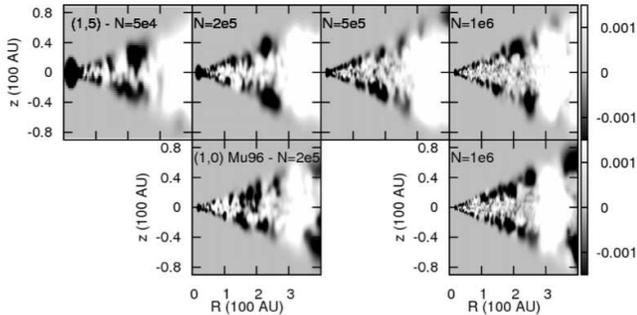}
\caption{\emph{Radial velocity structure of the gaseous disc:
changing resolution and artificial viscosity implementation}. 
The layout is the same as in Fig.~\ref{figNew-02}.
The top row shows how 
the radial velocity structure changes with resolution for simulations
with the \citetalias{Monaghan1983} artificial viscosity,
the specific case of simulations with ($\alpha$,$\beta$)=(1,5)
has been chosen as an example:
panels from left to right correspond respectively to
$N$=$5 \cdot 10^4$, $2 \cdot 10^5$, $5 \cdot 10^5$, $1 \cdot 10^6$.
The bottom row refers to two simulations with
the \citetalias{Murray1996} artificial viscosity implementation (with $\alpha=1$)
and with $N$=$2 \cdot 10^5$ (left panel) and $N$=$1 \cdot 10^6$
(right panel).
}
\label{figNew-03}
\end{figure}

Maps of the azimuthally averaged radial velocity
$\langle v_R \rangle_{\theta}$ in the $R$-$z$ plane 
are displayed in Figs.~\ref{figNew-02}~and~\ref{figNew-03} 
 for changing artificial viscosity and resolution respectively.

\underline{Effect of artificial viscosity parameters}.
A transition from a chaotic to a regular accretion pattern is clearly observed
in Fig.~\ref{figNew-02} when $\alpha$ is increased for a given $\beta$ 
(panels from top to bottom) and as well as when  
$\beta$ is increased for a given $\alpha$
(panels from left to right).

At large radii, a well extended decretion region
dominates in all models,
as expected from the conservation of angular momentum.
At smaller radii, instead, the structure of the gas flow  
strongly depends on the values of the AV parameters.
A decretion flow located around the midplane,
associated with an accretion flow 
in thin surface layers, appears when $(\alpha,\beta)$ are increased.
Such accretion layers become thicker for 
larger AV parameters.
Therefore, the gas flow  is characterised by 
the phenomenon of \emph{meridional circulation}.
This mechanism has been found
 in some two-dimensional viscous disc models 
(e.g. \citetalias{Takeuchi2002} that extends the standard
one dimensional \citetalias{Shakura1973} models 
to two dimensions: radial and vertical) but is not reproduced
 by MHD disc simulations where turbulence is induced by the 
MRI \citep{Fromang2011,Jacquet2013}.
The presence of meridional circulation has important implications for the topic
of particle mixing in PPD and is one of the possible mechanisms able to
explain the presence of crystalline  solid particles in the outer regions
of T~Tauri stars
recently observed by Spitzer \citep[e.g.][]{Sargent2009}.
Finally, in the very inner region of the disc a small and 
well defined accretion region is present in all models.

\underline{Effect of artificial viscosity implementation}. 
The two panels in the bottom row of Fig.~\ref{figNew-03} 
show the structure of the radial velocity flow of two
discs at different resolution simulated with the \citetalias{Murray1996} artificial viscosity.
The resulting flow structure can be directly compared with the 
panel above, which represents the flow structure 
in an equivalent disc at the same resolution but
simulated  with the \citetalias{Monaghan1983} artificial viscosity.
In the simulations at lower resolution ($N=2 \cdot 10^5$), 
the meridional circulation pattern is better reproduced by the \citetalias{Monaghan1983}
implementation.
However, in the more resolved discs ($N= 10^6$) the two 
patterns are very similar.
The two small accretion regions present in the outer \citetalias{Murray1996} discs
are due to a longer relaxation time-scale for this particular
artificial viscosity implementation 
(in longer evolution simulations, not shown here, 
we have observed a net outward gas flow 
in agreement with the \citetalias{Monaghan1983} case and the expected viscous spreading).

\underline{Effect of resolution}. 
The first row in  Fig.~\ref{figNew-03} shows how the radial
velocity structure has already converged after $N \approx 5 \cdot 10^5$
particles.

\citetalias{Takeuchi2002} found that the outflow zone around the midplane
shrinks as the radial volume density gradient is reduced
($s \rightarrow 0$) and when the condition
 $p+q<2$, that guarantees a net accretion 
of the gas onto the star, is violated.
Therefore, we expect that the thickness of the accretion layers 
depends on the values ($p$, $q$) of the surface density
and temperature radial profiles of the gaseous disc model.
However, a detailed study of the dependence 
of the structure of the accretion layers
on the gas density and temperature profiles 
and the correlation with the central mass accretion rate 
is out of the scope of this
paper and will be addressed in a future work.
For the initial profile of the disc model used in this work
 we have $p+q=2.25$, which  implies net decretion but is 
very close to the limit between net accretion and decretion.
The surface density of the simulated discs 
after numerical relaxation
reaches a smooth profile that, in the inner region, is slightly shallower 
(implying a slightly smaller $p$)
than the initially imposed power law, 
which is characterised by an inner sharp edge
(see Sect.~\ref{subsec:refdisc}).
This moves the model in the net accretion region and explains 
the presence of the observed thin accretion layers.

\subsubsection{Mach number of the flow}
The gas in the disc is supersonic.
In fact, since the gas moves on quasi-Keplerian orbits, the 
\emph{Mach number} is 
approximated by Ma~$\approx v_\mathrm{k}/c_\mathrm{s}$. With the adopted sound speed profile
it becomes: Ma~$\approx  R^{(q-1)/2}/c_{\mathrm{s}0}$.
For the values used here \mbox{Ma~$\approx 20 R^{-1/8}$}, which is 
approximately in the range of values 15--25.

\subsection{Mass accretion rate}
\label{sec:massAccretion}
The mass accreted onto the central star (Fig.~\ref{fig2})
and the radial profiles of the mass accretion rate
(Figs.~\ref{fig2c} and \ref{fig2c-2})
are two macroscopic signatures of turbulence.
\subsubsection{Central mass accretion rate}
\label{Subsubsec:centralMdot}
\begin{figure}
\centering
\includegraphics[width=0.5\textwidth]{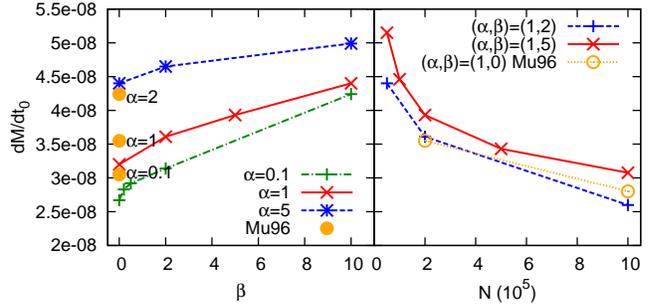}
\caption{\emph{Central mass accretion rate $\dot{M}_0$}. 
Left panel: dependence on the AV parameters 
for a given number of particles ($N=2 \cdot 10^5$),
Right panel: dependence on resolution.
The mass accretion rate is displayed in units of $M_\odot \textrm{yr}^{-1}$.
}
\label{fig2}
\end{figure}
\underline{Effect of artificial viscosity}.
The left panel in Fig.~\ref{fig2} shows that
  $\dot{M}_0$ increases with larger $\alpha$ and/or larger $\beta$
and in the case of the \citetalias{Murray1996} artificial viscosity implementation.
The effect of $\alpha$ is larger, as expected since it controls
the first order in the artificial viscosity term.

\underline{Effect of resolution}. 
The right panel in Fig.~\ref{fig2} shows that
 $\dot{M}_0$ decreases with increasing resolution.
Such an effect is due to a decrease of particle noise
at higher resolution, which decreases the numerical dissipation.
This is linked to the dependence of the effective viscosity
on resolution (see Sect.~\ref{subsec:effalpha}).
In fact, as will be shown in Sect.~\ref{subsubsec:PDFs},
the PDFs of several physical quantities
are narrower at higher resolution, due to the reduced noise.
Concerning the central mass accretion rate
we look at the part of the radial velocity distribution
corresponding to  \emph{negative} velocities,
since only fluid elements with negative radial velocity  
are responsible for accretion.
The radial velocity distribution at higher resolution 
presents a smaller spread  than in the  
lower resolution case.
In addition, SPH particles in low resolution simulations
have a larger mass than in higher resolution simulations.
Therefore, at lower resolution there is a larger 
number of more massive SPH particles with higher inward radial
velocity, implying higher accretion rates.
The trend of the curve shows that the central 
accretion rate is converging, 
however convergence is not yet reached at
one million particles.

For all the considered combinations of ($\alpha$,$\beta$) and $N$,
we find mass accretion rates consistent with the values observed for PPD.
In addition, we observe a correlation between the 
 trend of $\dot{M}_0$ with respect to $(\alpha$, $\beta)$ 
and the increase of thickness of the accreting layers
(net accreting mass flux)
observed in Fig.~\ref{figNew-02}. 
Similarly, the trend with respect to
$N$ correlates  with the decrease of thickness of the accreting layers
at higher resolution, observed in Fig.~\ref{figNew-03}.

\subsubsection{Radial profile of the mass accretion rate}

\begin{figure}
\centering
\includegraphics[width=0.45\textwidth]{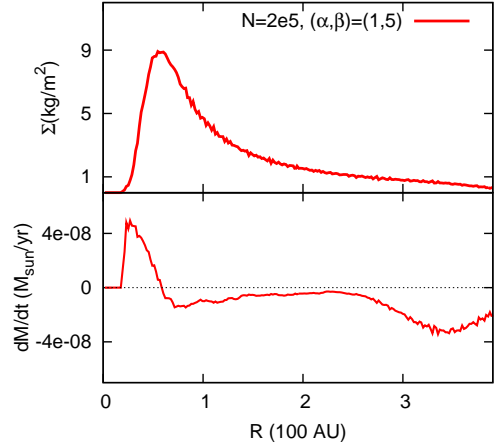}
\caption{\emph{Gas flow in the disc.} 
Surface density radial profile (top panel) 
and  mass accretion rate radial profile (bottom panel) 
for simulation S8.
}
\label{fig2c}
\end{figure}

The central mass accretion rate $\dot{M}_0$ gives information 
only about the region of the disc that is close to the star.
In order to have a global picture of the mass flux in the disc, it is 
interesting to look at the radial profile of the 
vertically integrated mass accretion rate $\dot{M}(R)$.
As an example, the profile observed in simulation S8 is 
displayed in the bottom panel of  Fig.~\ref{fig2c}.
The flow is characterised by an inner accreting region separated, at a
radius we call $R_t$, from an outer decreting region with small rates at
intermediate radii and larger rates at large radii.
It can be understood by a 
comparison to the surface density profile (top panel of Fig.~\ref{fig2c}).
In the organized flow of model S8, $R_t$ corresponds to the radial location
of the surface density peak, which moves outwards with time due to viscous
spreading, as the density maximum decreases. 
The density peak behaves as a reservoir of gas that supplies mass 
both to the inwards and outwards flows, resulting in an apparent
null mass accretion rate at $R_t$.
The varying rate in the decretion region is due to the vertical structure
of the flow: Fig.~\ref{figNew-02} shows accretion throughout the disc height
for the inner disc, accretion in the midplane with decretion in the
surface layers at intermediate radii resulting in a slightly negative
radial mass flux, and decretion throughout the disc height in the very
outer regions.

\begin{figure}
\centering
\includegraphics[width=0.45\textwidth]{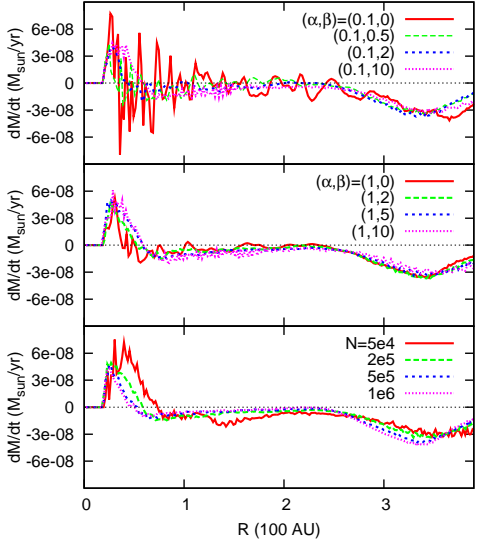}
\caption{\emph{Radial profiles of the mass accretion rate.}
The effect of artificial viscosity is shown 
for $\alpha=0.1$ (top panel) and 1 (middle panel) and changing $\beta$,
 for simulations with $N=2 \cdot 10^5 $ particles.
The effect of resolution is displayed in the bottom panel
 for simulations with $(\alpha, \beta)=(1,5)$.
}
\label{fig2c-2}
\end{figure}

\underline{Effect of artificial viscosity}.
For low values of $\alpha$ and $\beta$ fluctuations are so large that the
identification of an organized flow of the fluid is not possible.
As shown by the top panel of Fig.~\ref{fig2c-2}, the mass accretion rate
profile is very noisy with a narrow peak in the inner disc, an average value
close to zero in the intermediate regions and negative in the outer disc.  
The peak of the surface density profile decreases with time but less than
in the higher $(\alpha,\beta)$ case and it does not change position with time.
In the $v_R$ map (see the top left corner of Fig.~\ref{figNew-02}) 
negative and positive regions are clearly mixed.

Increasing the two AV parameters, the profile becomes more regular and smooth,
following the structure described above for the specific simulation S8. 
For $\alpha$~=~$1$ the profile tends to converge when $\beta$ is increased
(see the middle panel).
In the \citetalias{Murray1996} case we observe similar radial profiles for the mass accretion rate (not shown).

\underline{Effect of resolution}.
As shown in the bottom panel of Fig.~\ref{fig2c-2}
for the simulations with $(\alpha,\beta)=(1,5)$ and 
increasing resolution,
the peak of the radial profile of the mass accretion rate 
moves inwards and  converges for
$N \ga 5 \cdot 10^5$, in agreement with the convergence
observed for the radial velocity maps.
The reason of the decrease of the peak in higher resolution
simulations is the same that explains the 
similar decreases observed for the central mass accretion rate 
(see Sect.~\ref{Subsubsec:centralMdot}).

\subsection{The effective viscosity of the SPH disc: $\alpha_{\mathrm{\scriptstyle{2D}}}$}
\label{subsec:effalpha}
\begin{figure}
\centering
\includegraphics[angle=-90,width=0.45\textwidth]{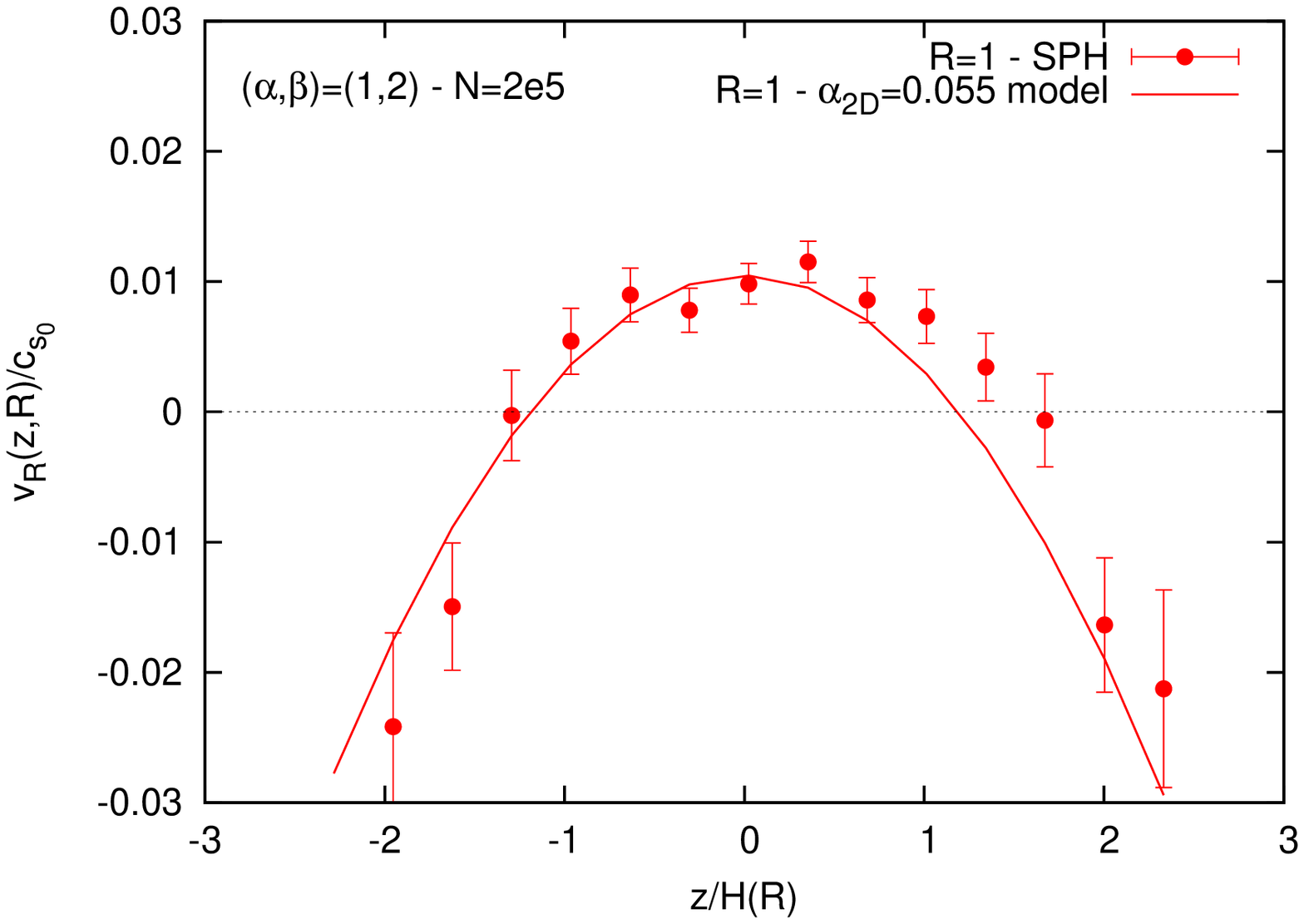}\\
\includegraphics[angle=-90,width=0.45\textwidth]{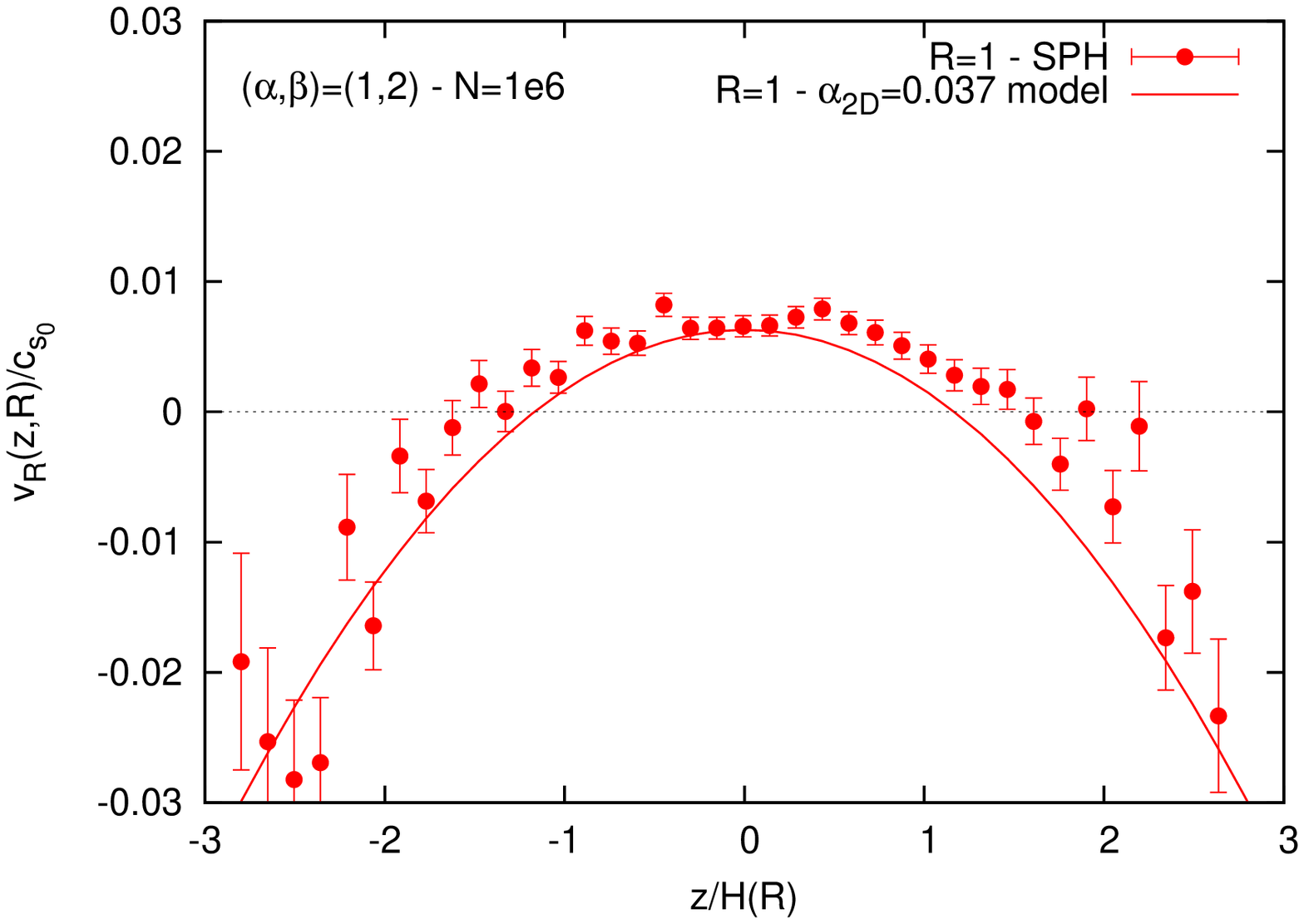}
\caption{\emph{Determination of $\alpha_{\mathrm{\scriptstyle{2D}}}$.}
The vertical profile of the radial velocity 
extracted from simulations  (points with error bars)
is fitted by the two-dimensional 
$\alpha_{\mathrm{\scriptstyle{SS}}}$-disc model (solid lines).
Top: simulation S7, bottom: simulation S20.
}
\label{fig4dd}
\end{figure}

In order to estimate the effective viscosity of the gaseous disc, 
we determine the  $\alpha_{\mathrm{\scriptstyle{2D}}}$ parameter,
which is  derived by fitting the vertical profile of the radial velocity
expected for a viscous axisymmetric accretion disc,
given by Eq.~\ref{eq:vr}, to that derived from the simulations
(see Appendix~\ref{app:fit} for the fitting procedure). 
Two examples, characterised by adequately large AV parameters,
 are shown in Fig.~\ref{fig4dd}, where
the vertical profile of the radial velocity  
at radial position $R=1$ 
is displayed for simulation S7 (top panel) and
for simulation S20 (bottom panel).

\begin{figure}
\centering
\includegraphics[width=0.5\textwidth]{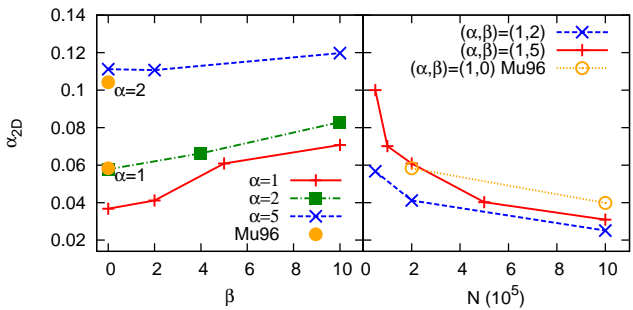}
\caption{\emph{Results for $\alpha_{\mathrm{\scriptstyle{2D}}}$.} 
The parameter has been determined 
by averaging the values fitted at $R$=1 and 2
and only for simulations 
where the noise is low enough to allow the fit.
}
\label{fig4d}
\end{figure}

In Fig.~\ref{fig4d} we present the values of 
$\alpha_{\mathrm{\scriptstyle{2D}}}$ obtained by 
averaging the result of the fit at $R$=1 and $R$=2
for the set of simulated discs.
The values are averaged in the radial range $[1,2]$.
In simulations with $\alpha=0.1$  
the vertical profile of radial velocity  cannot be described
by the profile in Eq.~\ref{eq:vr} because the high 
 noise level makes the fit impossible.
We find that simulation profiles can be fitted by a value of
$\alpha_{\mathrm{\scriptstyle{2D}}}$ which is roughly 
the same for the different
radial locations we considered.
This shows that the structure of SPH models is consistent
with the 2D version of \citetalias{Shakura1973} accretion disc models,
characterised by a constant $\alpha_{\mathrm{\scriptstyle{ss}}}$ value.

\underline{Effect of artificial viscosity}.
In the range of AV parameters were the determinations
of the effective viscosity  is possible ($\alpha>0.1$),
the scaling of $\alpha_{\mathrm{\scriptstyle{2D}}}$ with the AV parameters
(increase with larger $\alpha$ and/or $\beta$ in the \citetalias{Monaghan1983} implementation and 
increase with larger $\alpha$ for the \citetalias{Murray1996} implementation) has
the same qualitative behaviour as for the central mass accretion rate
(see Fig.~\ref{fig2}).
Such a match is what we expect, 
since a more viscous disc is characterised 
by a larger central accretion rate.
The weaker dependence on $\beta$ is in agreement 
with the fact that $\beta$ is the parameter controlling
the second order term of the \citetalias{Monaghan1983} artificial viscosity
(see Eq.~\ref{eq:AVMG83}).

\underline{Effect of resolution}.
The trend of the curves in the right panel of Fig.~\ref{fig4d}
is qualitatively similar to that of 
the central mass accretion rate (see Fig.~\ref{fig2})
and shows that the effective viscosity is converging.
However one million particles is still not enough
in order to reach convergence.
The slightly higher resolution required for the convergence
of  $\alpha_{\mathrm{\scriptstyle{2D}}}$ and $\dot{M}_0$
with respect to that required by the radial velocity maps and by
the mass accretion rate radial profile ($N \approx 5 \cdot 10^5$)
is due to the fact that the former two parameters are calculated 
from a smaller subset of simulation particles,
the numerical noise is therefore larger.

The result of this subsection is that the effective viscosity
$\alpha_{\mathrm{\scriptstyle{2D}}}$ in all the SPH models considered here
equals a few $10^{-2}$, in agreement with the order of magnitude deduced
from observations.

\subsubsection{Reynolds number}
\label{subsec:Re}
\begin{figure}
\centering
\includegraphics[width=0.45\textwidth]{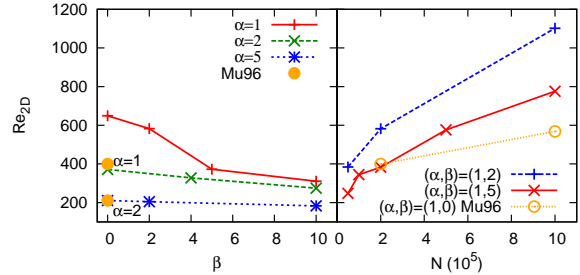}
\caption{\emph{The effective Reynolds number: $\mathrm{Re}_{2D}$.} Derived from the 
effective viscosity presented in Fig.~\ref{fig4d}.}
\label{figNew-04}
\end{figure}

Once the effective viscosity of the disc is determined
($\alpha_{\mathrm{\scriptstyle{eff}}} \approx \alpha_{\mathrm{\scriptstyle{2D}}}$),
it becomes possible to derive the corresponding Reynolds number
$\mathrm{Re}_{\mathrm{\scriptstyle{eff}}} \approx \mathrm{Re}_{\mathrm{\scriptstyle{2D}}} = 
\mathrm{Ma} / \alpha_{\mathrm{\scriptstyle{2D}}}$, shown in Fig.~\ref{figNew-04}.
In accordance with the trend of $\alpha_{\mathrm{\scriptstyle{2D}}}$ with 
respect to $\alpha$, $\beta$ and $N$, described in the last
paragraph, we found larger Re numbers in
 less viscous and more resolved discs.

Simulations with one million particles and the \citetalias{Monaghan1983} artificial
viscosity are closer to, but sill lower than the value  
$\mathrm{Re}_{\mathrm{\scriptstyle{eff}}} \approx 3 \cdot 10^3$, which is
considered as a limit for the onset of the turbulent regime
in a box with periodic boundaries 
\citep[see][]{Price2012}.
Therefore, the dynamics of the discs presented in this paper is not expected
to be dominated by fully-developed turbulence.

\subsubsection{The $\alpha_{\mathrm{\scriptstyle{2D}}}$-$\alpha_{\mathrm{\scriptstyle{cont}}}$ connection} 
\begin{figure}
\centering
\includegraphics[width=0.45\textwidth]{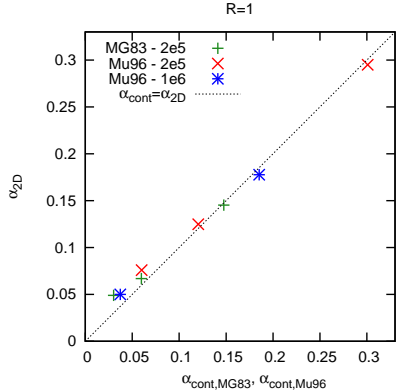}
\caption{
\emph{Comparison of the the effective viscosity of the disc $\alpha_{\mathrm{\scriptstyle{2D}}}$ with the corresponding $\alpha_{\mathrm{\scriptstyle{cont}}}$ formula}: simulations S6, S10 and S13 (\citetalias{Monaghan1983}, $N$=$2 \cdot 10^5$), 
S17, S18 and S27 (\citetalias{Murray1996}, $N$=$2 \cdot 10^5$) and S25 and S26 (\citetalias{Murray1996}, $N$=$ 10^6$).
The displayed values are computed at the radial position $R=1$. The dotted line represents identity between $\alpha_{\mathrm{\scriptstyle{cont}}}$ and $\alpha_{\mathrm{\scriptstyle{2D}}}$. 
Error bars  are comparable to the symbol sizes and are not displayed, their values are given in Table~\ref{table3}.
}
\label{fig4e}
\end{figure}

Figure~\ref{fig4e} shows the relation between the effective viscosity 
$\alpha_{\mathrm{\scriptstyle{2D}}}$ (computed in the simulations)
and the $\alpha_{\mathrm{\scriptstyle{cont}}}$ viscosity
(expected when the continuum limit of the SPH equations
is taken) for eight of the performed simulations
with $\beta=0$ (S6, S10, S13, S17, S18, S25, S27).
Simulations S1 and S16 (with $\alpha=0.1$) 
 are too noisy to allow the determination of $\alpha_{\mathrm{\scriptstyle{2D}}}$.
Since  the ratio $\langle h \rangle/H$ in the expression of
 $\alpha_{\mathrm{\scriptstyle{cont}}}$  depends on the radial location 
inside the disc (see Eqs.~\ref{eq:alphaSPH} and \ref{eq:alphaSPHMG83}),
we have evaluated locally at the radial position $R=1$ both $\alpha_{\mathrm{\scriptstyle{cont}}}$ and $\alpha_{\mathrm{\scriptstyle{2D}}}$, 
whose values are given in Table~\ref{table3}.

We find that both the points concerning the \citetalias{Murray1996}
case and those concerning the \citetalias{Monaghan1983} case follow 
the appropriate analytic relation.
We therefore confirm the necessity of reducing the 
coefficient in the $\alpha_{\mathrm{\scriptstyle{cont}}}$ 
relation in Eq.~\ref{eq:alphaSPHMG83} for the \citetalias{Monaghan1983}
case as shown by \cite{Meru2012} and motivated by the fact that in the
\citetalias{Monaghan1983} implementation the artificial viscosity is 
applied only to approaching particles.
Finally, no significant difference is observed when resolution is
changed from $2 \cdot 10^5$ to $10^6$ particles,
suggesting that already with $2 \cdot 10^5$ particles 
 the continuum approximation can be considered valid.

\begin{table}
\renewcommand{\arraystretch}{1.3}
\caption{Error bars associated to $\alpha_{2D}$ produced by the fitting procedure.} 
\label{table3}
\begin{center}
\begin{tabular}[]{l c c}
\hline
Simulation & $\alpha_{cont}$ & $\alpha_{2D}$ \\
\hline
S6 & 0.0301 & 0.0490 $\pm$  0.0063 \\
S10 & 0.0597& 0.0668 $\pm$ 0.0040\\
S13 & 0.1476& 0.1453 $\pm$ 0.0023\\
S17 & 0.0601& 0.0759 $\pm$ 0.0043\\
S18 & 0.1204& 0.1249 $\pm$ 0.0033\\
S27 & 0.3009& 0.2949 $\pm$ 0.0075\\
S25 & 0.0372& 0.0500 $\pm$ 0.0013\\
S26 & 0.1852& 0.1776 $\pm$ 0.0009\\
\hline
\end{tabular}
\end{center}
\end{table}

\section{The magnitude of SPH fluctuations}
\label{sec:magnoise}
\subsection{The Reynolds stress contribution to the SPH disc viscosity: 
$\alpha_{\mathrm{\scriptstyle{RS}}}$}
\label{sec:alphaRS}
We now proceed with the determination of $\alpha_{\mathrm{\scriptstyle{RS}}}$, which is the
$\alpha_{\mathrm{\scriptstyle{SS}}}$ parameter derived from the Reynolds stress,
in analogy with turbulence studies and with
 MHD or self-gravitating disc simulations.
Radial and azimuthal fluctuations are computed from the average velocity 
field determined locally using a $R$--$z$ grid.

\begin{figure}
\centering
\includegraphics[width=0.5\textwidth]{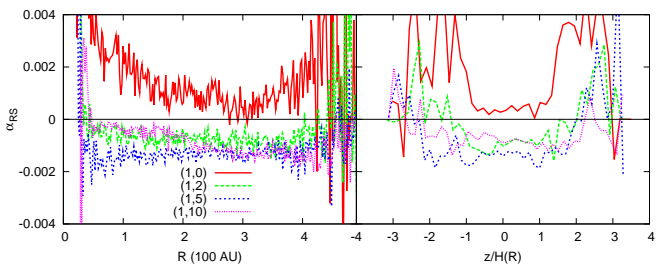}
\caption{\emph{Radial (left) and vertical (right) profiles of  $\alpha_{\mathrm{\scriptstyle{RS}}}$}.
Effect of changing the AV parameters.
}
\label{fig4}
\end{figure}

\begin{figure}
\centering
\includegraphics[width=0.5\textwidth]{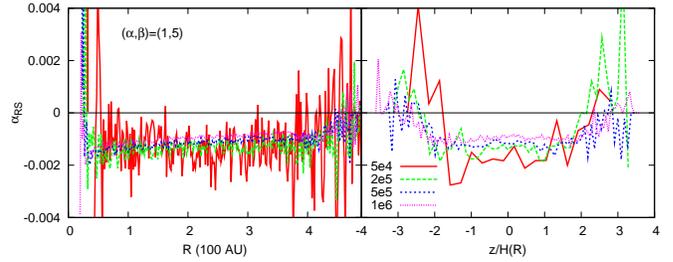}
\caption{\emph{Radial (left) and vertical (right) profiles of  $\alpha_{\mathrm{\scriptstyle{RS}}}$}.
Effect of changing the resolution.
}
\label{fig4-4}
\end{figure}

The radial profile of the vertically averaged $\alpha_{\mathrm{\scriptstyle{RS}}}$
is shown in the left panel of Fig.~\ref{fig4},
where the effect of changing $\alpha$ and $\beta$ is highlighted,
and of Fig.~\ref{fig4-4}, where the effect of resolution is shown.
In all simulations, the coefficient $\alpha_{\mathrm{\scriptstyle{RS}}}$  
is characterised by a similar profile: it tends to be approximately constant 
throughout the radial extension of the disc, 
with an increase both in the central and 
in the external radial region.
\begin{figure}
\centering
\includegraphics[angle=-90,width=0.45\textwidth]{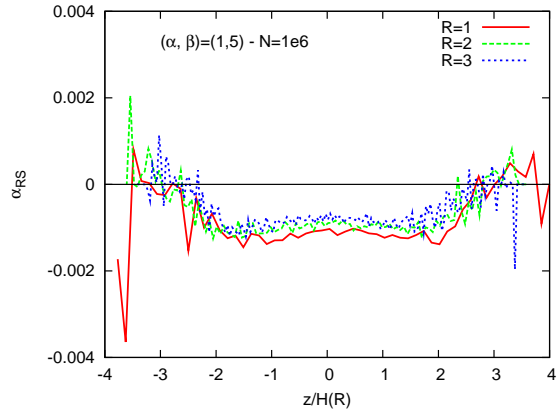}
\caption{\emph{Vertical profiles of  $\alpha_{\mathrm{\scriptstyle{RS}}}$}.
Comparison at three different radial location in the disc
for simulation S24.
}
\label{fig4cc}
\end{figure}

The right panel of both figures shows the radially averaged vertical profile
(as in Fig.~\ref{fig4dd}, the semi-thickness of the disc is computed from
$H(R)$=$H_0 (R/R_0)^{(3-q)/2}$, see Sect.~\ref{subsec:refdisc}):
$\alpha_{\mathrm{\scriptstyle{RS}}}$ is approximately constant 
around the midplane and then increases with height.
This behaviour qualitatively agrees with that observed
in MHD discs by \cite{Flock2011},
even if we observe a broader minimum.
In addition, we have found that 
the vertical profile is approximately the same at different radial positions,
as shown in Fig.~\ref{fig4cc} for simulation S24.

\underline{Effect of artificial viscosity}.
The radial and vertical profiles of  $\alpha_{\mathrm{\scriptstyle{RS}}}$
are very sensitive to the AV  parameters $\alpha$ and $\beta$.
In fact there is a sharp change in the average value, which decreases
from positive
to negative values, when the two AV parameters are
increased.
This transition is clearly visible in both panels of Fig.~\ref{fig4}
for $\alpha=1$: the increase of $\beta$ from 0 to 2 shifts the profile
of  $\alpha_{\mathrm{\scriptstyle{RS}}}$ 
towards smaller values, moving it in the negative region.
A further increase of $\beta$ leads to smaller changes,
but the behaviour inverts:
higher artificial viscosity now tends to 
increase the value of the coefficient $\alpha_{\mathrm{\scriptstyle{RS}}}$ 
(which corresponds to a decrease of its absolute value, see left panel)
and to extend the constant region around the midplane (see right panel).
We find the same trend for other values of $\alpha$ (not shown).

The change of sign of  $\alpha_{\mathrm{\scriptstyle{RS}}}$ is due to the
mechanism of meridional circulation that is correctly resolved when
adequately high AV parameters are used.

\underline{Effect of resolution}.
For a given combination of AV parameters,
the radial and vertical profiles
 of the $\alpha_{\mathrm{\scriptstyle{RS}}}$ coefficient
  do not change with resolution.
The only effect of increasing the number of particles
of the simulation is a reduction of the noise,
which implies smoother profiles.

We note that the computation of $\alpha_{\mathrm{\scriptstyle{RS}}}$ is very sensitive 
to the correct determination
of the average velocity field of the flow, 
therefore only in simulations with adequately high values of $\alpha$ and $\beta$, 
where numerical noise and particle disorder are lower, can it be trusted. 
The absolute value of $\alpha_{\mathrm{\scriptstyle{RS}}}$ 
for the present simulations is 
$|\alpha_{\mathrm{\scriptstyle{RS}}}| \approx 10^{-3}$.
This is comparable to standard
values found in  MHD discs \citep[see e.g.][]{Fromang2006}.
The difference is that meridional circulation is present here and the negative value indicates decretion on average in the midplane, in contrast to the positive value of MHD discs (where meridional circulation is not seen), which is expected for accretion in the midplane.

In conclusion, more viscous discs are characterised by a smaller modulus of 
$\alpha_{\mathrm{\scriptstyle{RS}}}$ and,
for all the simulations we performed, we find the ratio
of the Reynolds stress to the effective viscosity of the
gaseous disc to be
$\alpha_{\mathrm{\scriptstyle{RS}}}/\alpha_{\mathrm{\scriptstyle{2D}}}$~$\approx$~$10^{-1}$.
This means that velocity fluctuations are not the main component
of the effective viscosity of the disc, in agreement with the low
Reynolds number of the global flow in
in the present simulations (see Sect.~\ref{subsec:Re}).
However, the \emph{turbulent} Reynolds number,
which is associated to the Reynolds stress, amounts to
$\mathrm{Re}_{\mathrm{\scriptstyle{T}}}\approx 10^4$ (since Ma~$\approx 10$
and $\alpha_{\mathrm{\scriptstyle{RS}}} \approx 10^{-3}$),
close to the turbulent limit.
Therefore, physical fluctuations, quantified by the small contribution of
$\alpha_{\mathrm{\scriptstyle{RS}}}$ to the effective viscosity,
can be expected to exhibit turbulent signatures. We will show in
Sects.~\ref{subsec:diff} and \ref{subsec:powerspectrum} that it is indeed
the case.

An important point to be highlighted is that 
the determination of $\alpha_{\mathrm{\scriptstyle{RS}}}$
is not a sufficient condition in order to determine the
mass accretion rate of the disc, since other sources 
can be present in the model.

\subsection{Diffusion coefficient}
\label{subsec:diff}
\begin{figure}
\centering
\includegraphics[width=0.23\textwidth]{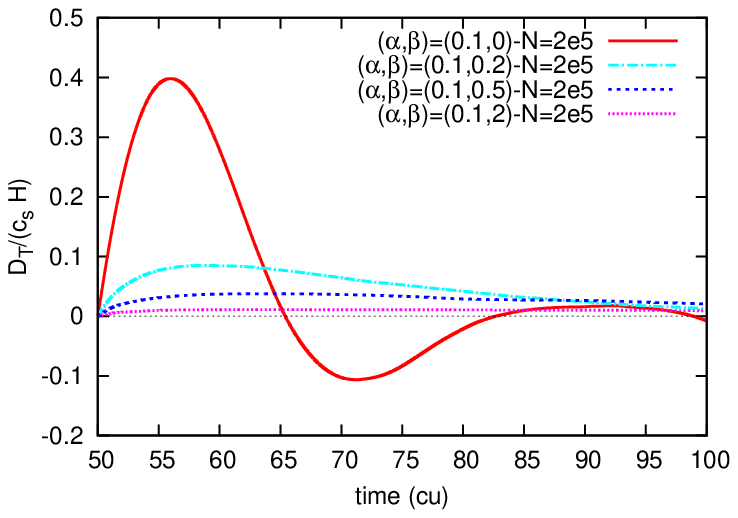}
\includegraphics[width=0.23\textwidth]{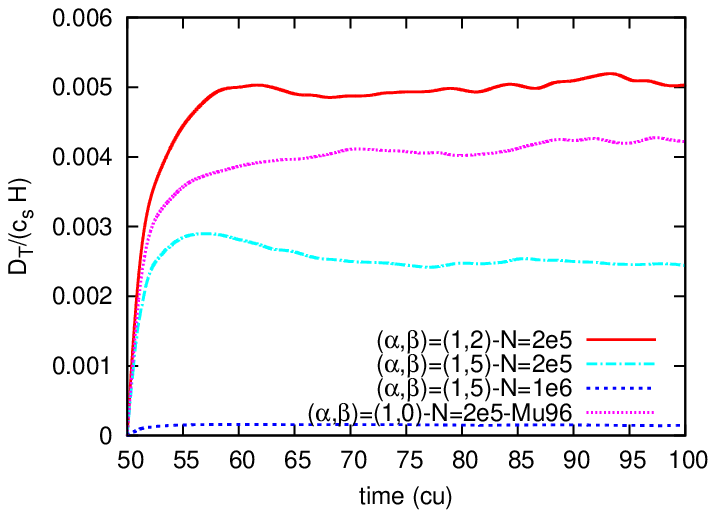}
\caption{\emph{Time evolution of the diffusion coefficient.}
$D_{\mathrm{\scriptstyle{T}}}/(c_\mathrm{s} H)$ computed at $R=2$.
Left: low artificial viscosity simulations (S1, S2, S3, S4).
Right: higher artificial viscosity (S7, S8, S24
with \citetalias{Monaghan1983} implementation and S17 with \citetalias{Murray1996} implementation).
Note the different vertical scales.
}
\label{fig6}
\end{figure}

For each simulation, 
we have computed the diffusion coefficient from Eq.~\ref{eq:diffcoeff}
at several radial positions
in the disc. In Fig.~\ref{fig6} we show 
the evolution with time of the diffusion coefficient $D_{\mathrm{\scriptstyle{T}}}$,
 in units of $c_\mathrm{s} H$, measured at the intermediate radial position 
$R=2$ for some simulations belonging to Set A.
Except for simulations from S1 to S3, characterised by larger fluctuations,
the diffusion coefficient regularly increases with time and converges to 
a well-defined value.

\begin{figure}
\centering
\includegraphics[width=0.5\textwidth]{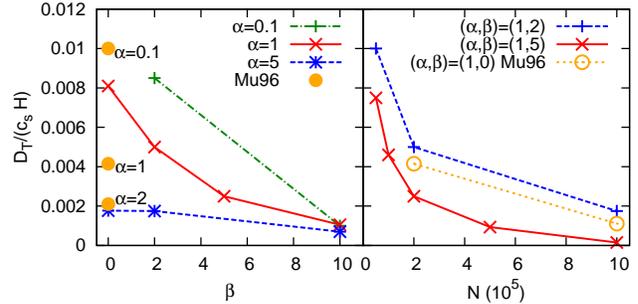}
\caption{\emph{Results for $D_{\mathrm{\scriptstyle{T}}}$.}
Left: effect of artificial viscosity, right: effect of resolution.
}
\label{fig6b}
\end{figure}

For the standard combinations of parameters $(\alpha,\beta)=(1,2)$,
$D_{\mathrm{\scriptstyle{T}}}/(c_\mathrm{s}H) \approx 5 \cdot 10^{-3}$ 
is in agreement with 
the value found by \cite{Fromang2006a} 
in their MHD local shearing box stratified simulations.
The asymptotic values of the diffusion coefficient for
each performed simulation are displayed in Fig.~\ref{fig6b}.

\underline{Effect of artificial viscosity}.
The left panel of Fig.~\ref{fig6b} shows that the coefficient 
decreases with increasing $\alpha$ and $\beta$
(except simulations from S1 to S3 which present a diffusion
coefficient without a clear trend, due to the higher noise level)
and in the \citetalias{Murray1996} case (the artificial viscosity switch 
on approaching/receding particles is off, resulting in more viscous discs).
This trend is due to a reduction of the noise that allows 
the onset of meridional circulation,
which forces the fluid to be more localised in the vertical direction.

\underline{Effect of resolution}.
From the right panel of Fig.\ref{fig6b} we observe 
that the diffusion coefficient 
decreases with increasing resolution.
The reason of this behaviour is the same that
explains the qualitatively similar trend 
we have observed in Sect.~\ref{Subsubsec:centralMdot} for the central 
mass accretion rate.
In this case, the PDF of the vertical velocity
$v_z$  is broader
in low resolution simulations
with respect to more resolved simulations.
Therefore, larger vertical velocity are possible
in less resolved discs with resulting 
 higher values (see Eq.~\ref{eq:vz}) of the diffusion coefficient.
The curve shows a converging trend, however
convergence is not yet reached at one million particles.
As for the central mass accretion rate and for the effective viscosity fit,
convergence requires more resolution because the diffusion coefficient is
computed from a smaller subset of particles than 
in the case of the radial velocity maps and the mass accretion rate profiles,
where an average on the azimuthal component is taken.

We conclude that the diffusive mechanism is correctly 
represented by intermediate/large AV parameters $(\alpha, \beta)$:  more viscous discs are less diffusive, due to the onset of meridional circulation
and the decrease of $\alpha_{\mathrm{\scriptstyle{RS}}}$.
In addition, for a given $(\alpha,\beta)$ combination,
less resolved discs are more diffusive:
such numerical dependence is of the same type as the dependence of the central mass accretion rate and of the effective viscosity on the resolution of the simulation (see Sect.~\ref{Subsubsec:centralMdot}.) 

\subsection{Mach number of velocity fluctuations}
\label{sec:mach}
\begin{figure}
\centering
\includegraphics[width=0.24\textwidth]{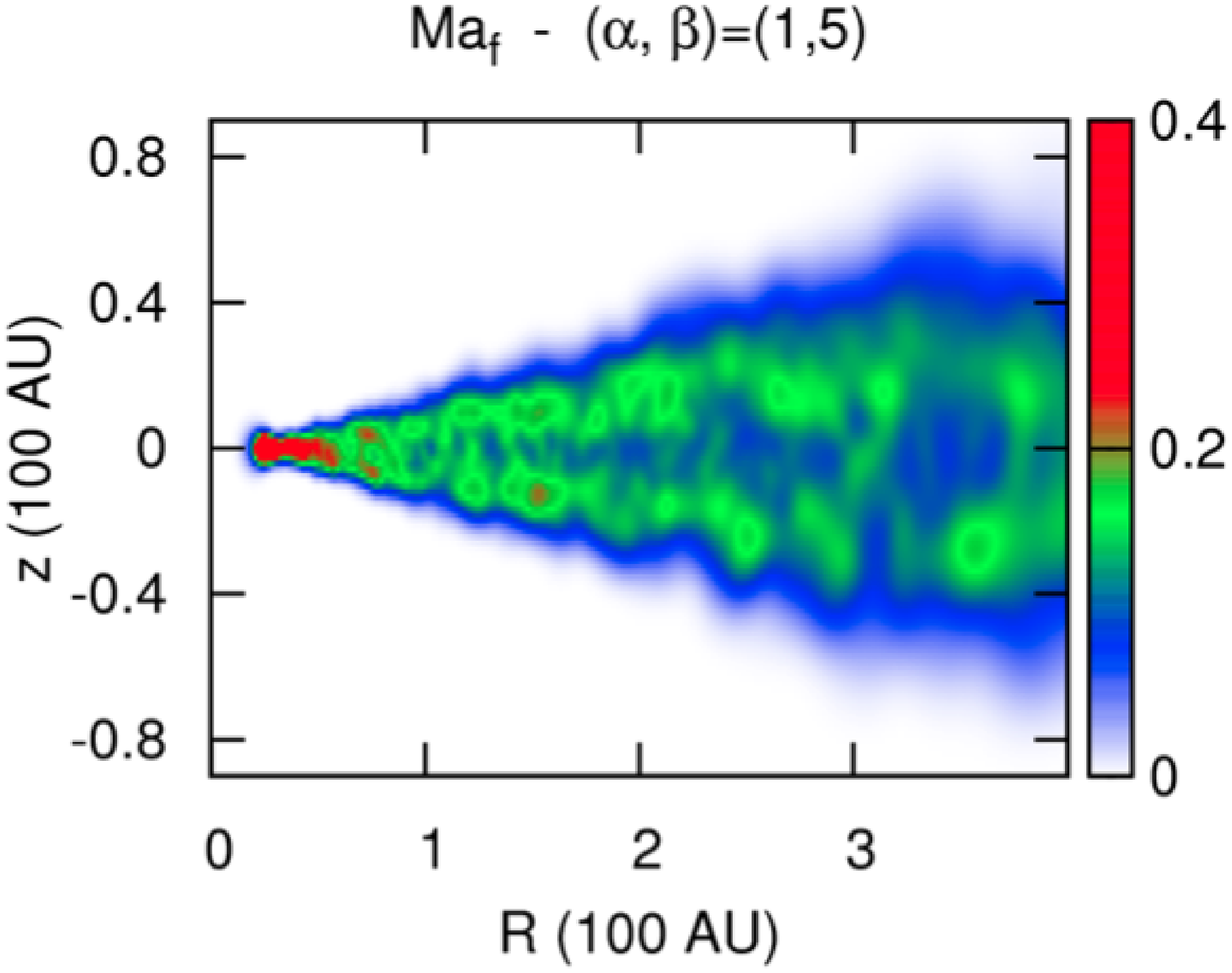}
\includegraphics[width=0.23\textwidth]{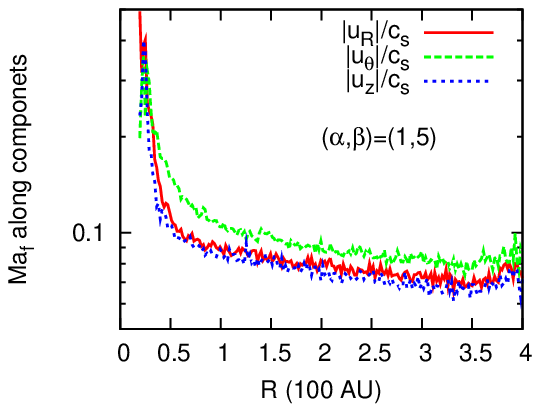}
\caption{\emph{Mach number of velocity fluctuations.}
Left: azimuthal average in the  $R$--$z$ plane,
right: radial profile of the vertical and azimuthal average for each velocity
component, in simulation S23. 
}
\label{fig6e}
\end{figure}

The azimuthally averaged Mach number of the modulus of velocity fluctuations 
is displayed in the $R$--$z$ plane in the left panel of Fig.~\ref{fig6e}
for simulation S23, as an example.
The right panel shows the corresponding radial profile for each
velocity component, vertically and azimuthally averaged.
All simulations show qualitatively similar structures and profiles.

\begin{figure}
\centering
\includegraphics[width=0.5\textwidth]{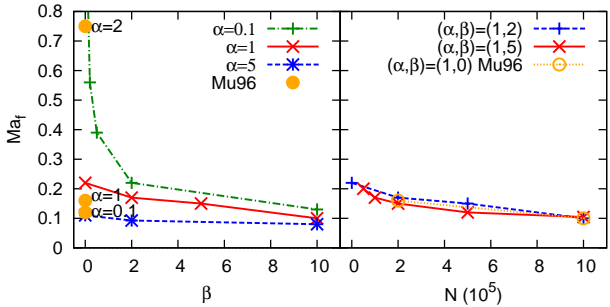}
\caption{\emph{Results for $\mathrm{Ma}_\mathrm{f}$.} 
Effect of changing the AV parameters and the AV switch (left panel) 
and resolution (right panel).}
\label{fig6d-2}
\end{figure}

\underline{Effect of artificial viscosity}.
The average value of $\mathrm{Ma}_\mathrm{f}$ in the radial range $R \in [1,2.5]$ is displayed 
in Fig.~\ref{fig6d-2}.
From the left panel of the figure,
we see that, except for simulation S1, 
whose fluctuations are close to the sonic regime,
all simulations show subsonic fluctuations with  Mach number
significantly smaller than unity.
We observe a decrease of the average Mach number with larger AV parameters.
For relatively high $\alpha$ and $\beta$, at each given radius the Mach number
increases from a minimum value in the midplane ($\mathrm{Ma}_\mathrm{f} \approx 0.1$) to 
a maximum value in the surface layers ($\mathrm{Ma}_\mathrm{f} \approx 0.2$),
as displayed by the left panel of Fig.~\ref{fig6e}.
This behaviour is qualitatively in agreement with what is found
in MHD simulations
\citep[e.g.][]{Fromang2006,Flock2011}, however we observe
smaller values.
A global maximum $\mathrm{Ma}_\mathrm{f} \approx 0.4$ is reached in the inner region, 
where $v_R < 0$ 
and the two accreting layers meet.
In addition, two layers with locally higher 
fluctuations with $\mathrm{Ma}_\mathrm{f} \approx 0.2$ are also present
at the interface
between the surface accreting layers and the midplane outwards flow.

\underline{Effect of resolution}.
Similarly, as a consequence of noise reduction, 
higher resolution leads to a decrease both of the Mach number and 
of the extension of the surface layers where it reaches the local maximum
for a given radial position (with the same qualitative behaviour
of the accreting layers in the $v_R$ maps, see
Figs.~\ref{figNew-02} and \ref{figNew-03}), as expected. 
We have found a convergence of the vertically and azimuthally averaged 
Mach number for a number of particles $N \approx 5 \cdot 10^5$ 
(see right panel of Fig.\ref{fig6d-2}), in agreement with 
the value found for radial velocity maps (which are connected to 
radial velocity fluctuations and therefore 
to the radial part of the Mach number).

Fluctuations tends to be dominant in the regions  
of the disc where accreting and decreting flows
meet: the cone-like regions between the accreting surface layers 
and the decreting midplane layer, and the 
region around the inner disc edge.

\section{The structure of SPH fluctuations}
\label{sec:strunoise}
\subsection{Anisotropy of velocity fluctuations}
\begin{figure}
\centering
\includegraphics[width=0.5\textwidth]{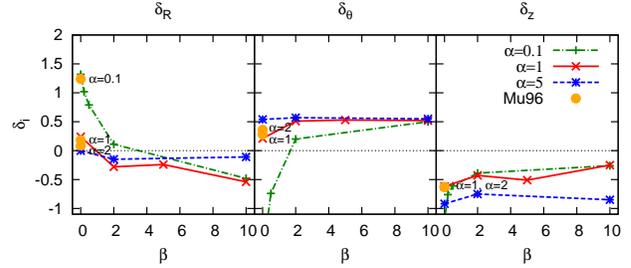}
\caption{\emph{Results for $\vec{\delta}$, 
the anisotropy vector of velocity fluctuations.} 
Effect of changing the AV parameters and the AV switch (\citetalias{Monaghan1983} and \citetalias{Murray1996}), values
are averaged vertically, azimuthally and in the radial domain $[1,2.5]$.
Note that values of the azimuthal and vertical anisotropy 
(middle and right panel) for $\beta=0$ become highly negative, they fall
outside of the displayed area.}
\label{fig7c-2}
\end{figure}

We recall that when  $\delta_R = 2$ the system is \emph{totally}
radially anisotropic, while for negative values
 the system is dominated by anisotropy in different directions.
The same holds for the azimuthal and vertical components.
The kind of anisotropy of the disc is displayed in 
Fig.~\ref{fig7c-2} for 
simulations with changing $\alpha$ and $\beta$.
We observe that none of the simulated discs is totally
anisotropic in one particular direction, but
there is always a component that dominates the other two.

\underline{Effect of artificial viscosity}.
Discs with smaller AV parameters
 are dominated by radial anisotropy.
For example, in the case of  simulation S1 with ($\alpha$,$\beta$)=(0.1,0),
we see from Fig.~\ref{fig7c-2} 
that $\delta_R \approx 1.4$, $\delta_{\theta} < -1$ and
$\delta_z < -1$, it follows that $\delta_R$
is significantly larger both than $\delta_{\theta}$ and
than $\delta_z$.
When $\alpha$ and/or $\beta$ are increased,
$\delta_R$ decreases and becomes negative
while $\delta_{\theta}$ increases.
The vertical anisotropy is always negligible 
with respect to the radial or azimuthal anisotropy
(it is always negative in all performed simulations,
as shown by the right panel in Fig.~\ref{fig7c-2}).

\underline{Effect of resolution}
More resolved simulations are slightly more 
azimuthally anisotropic.
As for some of the previously considered quantities,
we observe that the components of the anisotropy vector,
averaged in the vertical, azimuthal and in the usual radial range 
$R \in [1,2.5]$, converge at $N \approx 5 \cdot 10^5$ (not shown),
for the same reason already outlined several times.

We note that radial anisotropy has been observed in MHD discs
\citep[see e.g.][]{Fromang2006a,Fromang2006} where the mechanism
of meridional circulation is absent.
Here we observe radially dominated discs only for low
AV parameters, for which meridional circulation is also absent.

\subsection{Power spectrum}
\label{subsec:powerspectrum}
\begin{figure}
\centering
\includegraphics[width=0.5\textwidth]{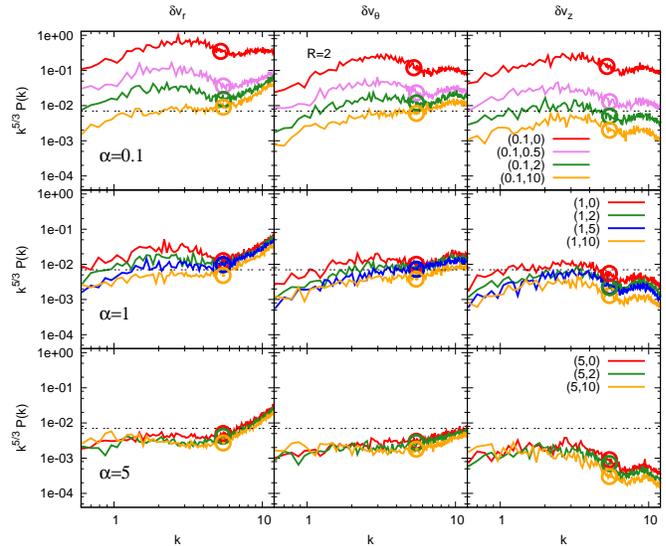}
\caption{\emph{Compensated power spectrum $k^{5/3}P(k)$ of the velocity field.}
Effect of changing $\alpha$ and $\beta$. 
The three columns from left to right refer respectively to 
radial,  azimuthal and vertical velocity components. 
The power spectrum is computed along a ring located at $R=2$ and $z=0$.
The circle on each curve marks the wavenumber 
corresponding to the scale of the smoothing length.
The number of particles is $N = 2 \cdot 10^5$.
}
\label{fig8}
\end{figure}

\begin{figure}
\centering
\includegraphics[width=0.5\textwidth]{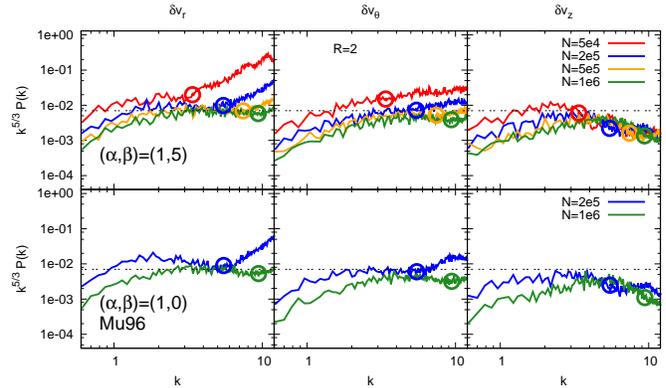}
\caption{\emph{Compensated power spectrum of the velocity field.}
Effect of changing  changing the AV switch and the resolution. 
The circle on each curve marks the wavenumber 
corresponding to the scale of the smoothing length.
}
\label{fig8b}
\end{figure}

For each simulation we have computed the power spectrum 
of velocity fluctuations in a ring located at different radii 
from the central star and at several altitudes, as explained
in Sect.~\ref{subsec:gtf}. As an example, in Figs.~\ref{fig8}~and~\ref{fig8b}
we present compensated spectra  
(divided by the Kolmogorov slope of $-5/3$, for easy comparison)
evaluated at the midplane location $(R,z)$=$(2,0)$
Different locations in the discs are considered later on.

Since we are dealing with an \emph{anisotropic} system,
we look at the one-dimensional power spectrum in the three directions:
 radial, azimuthal and vertical, which are 
given respectively in the first, second and third column
of each row of the two figures.

At least three regions can be identified 
in the power spectra extracted from simulations.
These regions are defined by two wavenumbers: $k_1 < k_2$.
The smallest scale corresponds to the resolution limit
of the simulation that is given by the smoothing length
at the considered position $h(R,z)$
(since the discs are axisymmetric there is not
dependence on $\theta$) and defines the largest 
wavenumber $k_2$ (for each curve, this is marked 
by a circle in the plots).
The region beyond $k_2$ corresponds to length scales inside the
smoothing length, which are below the resolution.
Therefore, the resolved region of the spectrum 
is to the left of the circle.
The other scale marks the boundary between
regions of the spectrum  described 
by power laws $P(k)=P_0 k^{a}$ with a different slope:
for $k<k_1$ the slope $a_1$ is close to zero,
for  $k_1<k<k_2$ the slope is called $a_2$ and is  
always negative.

In the case of isotropic turbulence one could identify 
$k_1$ with the `forcing scale' and the region between
$k_1$ and $k_2$ with the `inertial range'.
However, in the present anisotropic case, it is not possible
to find a direct link between $k_1$ and the forcing scale in the
disc because of the \emph{aliasing phenomenon} present in
one-dimensional spectra of turbulent shear-stress fluids
\citep[see e.g.][]{Pope2000}, 
since the power corresponding to wavenumber
$k_1$ also contains contribution from larger wavenumbers.

In the following we focus on the qualitative evolution 
of the power spectrum with changing artificial viscosity and resolution.

\underline{Effect of artificial viscosity}.
What is interesting is the dependence of the shape
of the power spectrum on the AV parameters.
Simulations with $\alpha \lesssim 1$ and $\beta \lesssim 2$ show
a positive slope of the compensated spectra
with a decrease just before the resolution scale
($k \lesssim k_2$), which mimics
the decay corresponding to the dissipation scale. 
In this case a  cascade is not clearly identified
in any of the three directions.
An example is given by  simulation S1 with 
($\alpha$,$\beta$)=(0.1,0) in the
top row of Fig.~\ref{fig8}.

For higher AV parameters 
the behaviour of the power spectrum depends on the
component of the velocity fluctuations,
confirming that we are dealing with an anisotropic system
also at intermediate scales.
In particular, 
for radial velocity fluctuations 
in the case of  $\alpha>1$ and $\beta>2$
(e.g. middle and bottom left panels of Fig.~\ref{fig8})
and for the azimuthal 
component when $\alpha \approx 5$ and $\beta>2$
(bottom central panel of Fig.~\ref{fig8}),
a progressively more extended 
flat region (indicating a power spectrum with a slope close to
the  Kolmogorov value)
is visible between the positive slope region on the left
and the decay immediately before the resolution scale
on the right.
The other azimuthal spectra show a positive slope
(e.g. top and middle central panels), indicating a sub-Kolmogorov slope. 
Concerning the vertical component of the velocity, 
flat regions start to appear for values of $\alpha$ close to 1, 
just above the resolution limit.
However, they are less extended than in the case of 
the other two components. 
For higher $\alpha$, we observe that larger $\beta$
 leads to a steeper slope.

In correspondence to the onset of meridional circulation,
we observe  the appearance of an approximately flat region
in the compensated spectra for all velocity components and 
at the smallest resolved scales.
This result suggests that the system is more isotropic 
at the small scales above the resolution limit.

\underline{Effect of resolution}
In agreement with  previous analyses, we observe 
a convergence of the power spectrum of velocity fluctuations
when a number of particles $N \approx 5 \cdot 10^5$ is used.
In addition, higher resolution  discs are characterised by a cascade  that is 
globally more extended in the wavenumber space than in lower resolution discs
because they resolve smaller length scales,
corresponding to larger wavenumbers.
The resolution scale $k_2$ therefore shifts towards larger  values
(see Fig.~\ref{fig8b}). 

In numerical simulations of turbulence 
a pile-up of energy (positive slope of the spectrum) is often observed
 in the high wavenumber region of the power spectrum 
(corresponding to the pre-dissipative region, close to grid resolution)
and is probably due to numerical dissipation.
We observe a similar feature only in the power spectrum of radial 
velocity fluctuations in intermediate and large viscosity discs
(see left panels in Fig.~\ref{fig8} and \ref{fig8b}).
The pile-up is reduced by increasing resolution
(see left panels in the top two rows in Fig.~\ref{fig8b}),
supporting the numerical origin of the phenomenon.

Simulations are observed  to converge towards a power spectrum
with a cascade close to the Kolmogorov one for 
the radial and azimuthal components.
The system reflects its anisotropy 
in a cascade that is always more extended for radial fluctuations 
and very short for vertical velocity fluctuations.
Such a particular feature of the cascade is due to the different
physical scales present in the disc, which is 
more radially than vertically extended (the disc is thin).

\underline{Changing location in the disc}.
The trends  of the power spectrum observed with changing 
 artificial viscosity and resolution  are qualitatively similar 
at different radial locations in the disc (not shown).
\begin{figure}
\centering
\includegraphics[width=0.5\textwidth]{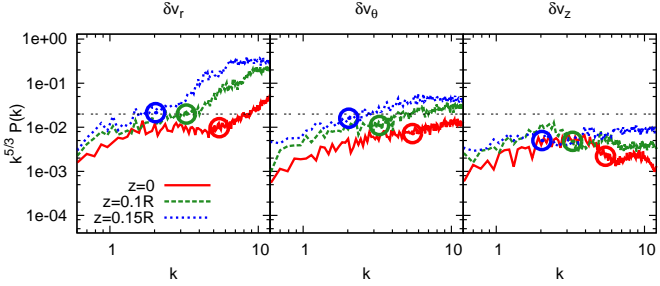}
\caption{\emph{Compensated power spectrum of the velocity field.}
Effect of changing the location inside the disc.
Power spectrum computed along a ring located at $R=2$  
and at three different heights above the midplane
for simulation S8 with $(\alpha,\beta)=(1,5)$.
}
\label{fig9}
\end{figure}
In Fig.~\ref{fig9} spectra at different heights above the midplane 
are compared for each of the three velocity components in the case of 
simulation S8 where $(\alpha,\beta)=(1,5)$: 
moving away from the midplane, 
the slope of the cascade becomes slightly smaller and 
the cascade disappears due to decreasing resolution.
In fact, 
at higher altitude the gas density is lower,
implying a larger smoothing length (i.e. worse resolution).
The wavenumber associated to the larger smoothing length
is smaller and therefore it is closer to the wavenumber of the
`forcing scale', shortening  the cascade.

We conclude that the power spectrum preserves its properties throughout the vertical and radial extension of the disc in the resolved regions of the wavenumber space.

\subsection{Intermittency}
\label{sec:inter}
In order to determine if intermittency is present in the simulations
we analysed the PDF of the 
density and of the acceleration in the azimuthal direction 
(the direction of the shear)  and the 
$3^{\mathrm{\scriptstyle{rd}}}$ and $4^{\mathrm{\scriptstyle{th}}}$ 
order moment of the PDF of several quantities.

\subsubsection{Density and azimuthal acceleration PDF}
\label{subsubsec:PDFs}
\begin{figure}
\centering
\includegraphics[width=0.5\textwidth]{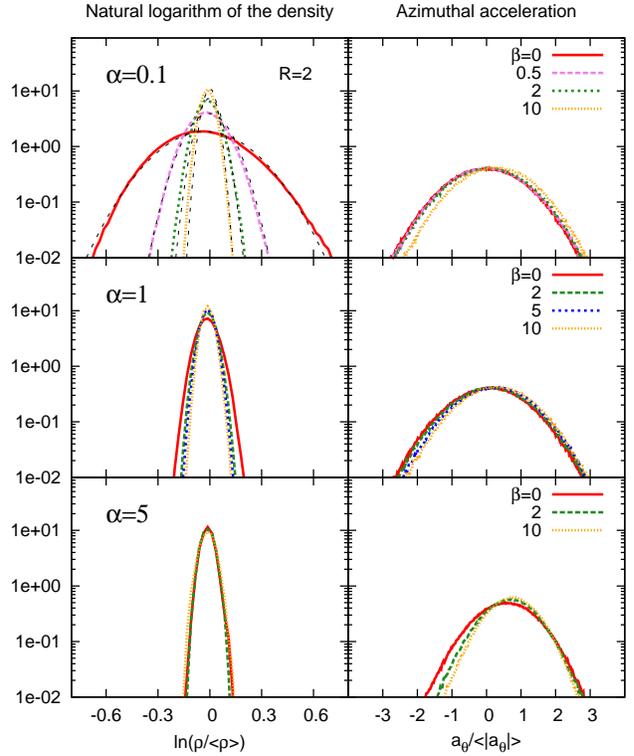}
\caption{\emph{Probability distribution functions.}
Effect of AV parameters.
Left: natural logarithm of the density. 
Right: acceleration.
PDFs are computed at the midplane and at distance $R=2$ 
from the central star.
The density PDF is compared to a log-normal distribution
(dashed line) in the top left panel.}
\label{fig11}
\end{figure}
\begin{figure}
\centering
\includegraphics[width=0.5\textwidth]{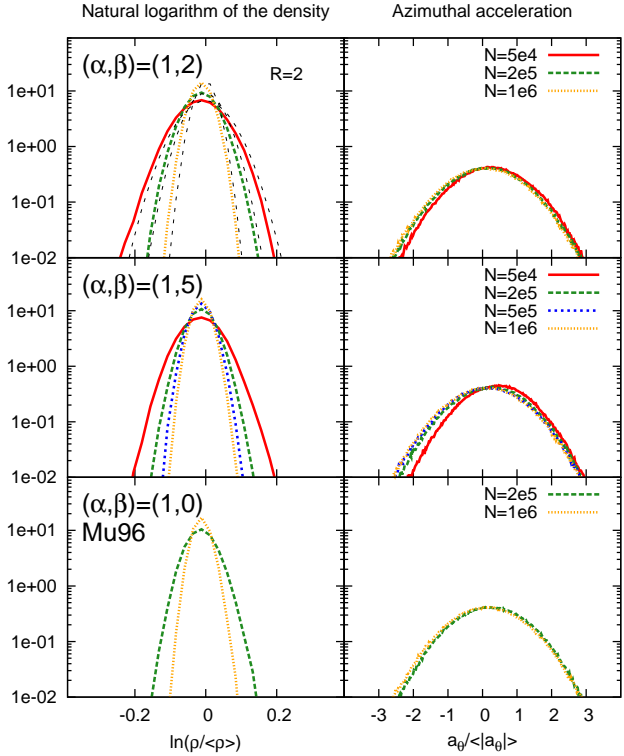}
\caption{\emph{Probability distribution functions.}
Effect of resolution.
Left: natural logarithm of the density. 
Right: acceleration.
PDFs are computed at the midplane and at distance $R=2$ 
from the central star.
The density PDF is compared to a log-normal distribution 
(dashed line) in the top left panel. }
\label{fig11b}
\end{figure}
\underline{Effect of artificial viscosity}.
The effects of the AV parameters
on the shape of the 
PDF of the natural logarithm of the density
 are shown in
the left panel of Fig.~\ref{fig11}.
The distribution of ln$(\rho/\langle \rho \rangle)$ 
is  described by a Gaussian distribution. 
It follows that the density is characterised by
a log-normal distribution, 
as observed in simulations of supersonic compressible turbulence
 \citep[see e.g.][and Sect.~\ref{sec:struct}]{Price2010}.

However, in our simulations the standard deviation of the
density logarithm $\sigma_p$, which controls the width of the distribution,
is much smaller, since we are dealing with subsonic 
physical fluctuations 
(as shown in Sect.~\ref{sec:mach}, Fig.~\ref{fig6e}
and Eq.~\ref{eq:sigmap}). In addition, we observe that 
larger values of $\alpha$ and/or $\beta$ lead to a narrower
distribution. This behaviour is explained by the same 
reason: the Mach number of physical fluctuations decreases 
and leads to a smaller $\sigma_p$, which in turn produces a PDF close to
a Gaussian distribution and gives values of the skewness and kurtosis
close to Gaussian values (as we will show in Sect.~\ref{subsec:SK}
and as expected from Eq.~\ref{eq:SK}).
The effect of changing $\beta$ for a given $\alpha$ is smaller
than that of changing $\alpha$ for a given $\beta$, due to the
fact that $\beta$ controls the second order term in the
\citetalias{Monaghan1983} artificial viscosity.

It should be noted here that even for high $\alpha$ and/or $\beta$,
although the effective viscosity becomes larger, 
the Reynolds stress is always non zero and still contributes to it.
Therefore physical fluctuations are present and the disc is not laminar.
For example, in our discs when $\alpha \approx 5$ 
the effective viscosity increases to 
$\alpha_{\mathrm{\scriptstyle{2D}}} \approx 0.1$ and
the Reynolds stress amounts to
$\alpha_{\mathrm{\scriptstyle{RS}}} \approx 10^{-3}$,
with a corresponding turbulent Reynolds number 
$\mathrm{Re}_{\mathrm{\scriptstyle{T}}}\approx 10^4$,
see Sect.~\ref{sec:alphaRS} and Figs.~\ref{fig4}~and~\ref{fig4-4}.
If the disc were laminar, one would not expect
a distribution of densities but a single value,
and therefore a delta function.

In the right column of Fig.~\ref{fig11} we show the PDF
of the acceleration in the 
\emph{azimuthal} direction since it presents
an interesting feature: the distribution shifts
towards higher values when 
both $\alpha$ and $\beta$ are increased.
This is an effect of the onset of meridional circulation
in discs with high enough artificial viscosity.
In fact, the presence of meridional circulation imposes
a net positive radial velocity (outward flow) to the gas
around the midplane that, combined with the disc rotation,
implies a spiral-like flow. The acceleration is thus
characterised by a positive average azimuthal component
(in contrast to a pure circular flow where 
the acceleration is only radial).
The peak location moves towards the value 
$a_{\theta}/\langle |a_{\theta}|  \rangle$=1
for larger AV parameters, 
however  it does not reach unity
 even for the larger AV combination
considered here ($\alpha, \beta$)=(5,10)
since a fraction of the particles still presents
a small but negative azimuthal component of the
acceleration.
Finally, the PDF of the radial and vertical components
of the acceleration (not shown here) present
the standard features of a pure circular flow:
both of them have a symmetric shape with the peak
located around the non-zero Keplerian value 
or around zero, respectively.

\underline{Effect of resolution}.
The effect of resolution is shown in Figure~\ref{fig11b}.
For all the combinations of AV parameters, 
more resolved discs are characterised by a narrower density
PDF (left column), due to the reduction in the numerical noise. 
The azimuthal acceleration PDF (right column) is only slightly affected,
with the peak shifted towards smaller acceleration for
higher resolution, due to a correspondingly lower strength
of  meridional circulation (see Sect.~\ref{subsec:velstruc}).
As already found in previous analysed quantities, 
a number of particles as large as $N \approx 5 \cdot 10^5$ 
guarantees a good degree of convergence.

In conclusion, the PDFs of density fluctuations reproduce those expected
for non intermittent subsonic turbulent flows, 
confirming that the discs are not laminar.

\subsubsection{Higher order moments: S and K}
\label{subsec:SK}
\begin{figure}
\centering
\includegraphics[width=0.55\textwidth]{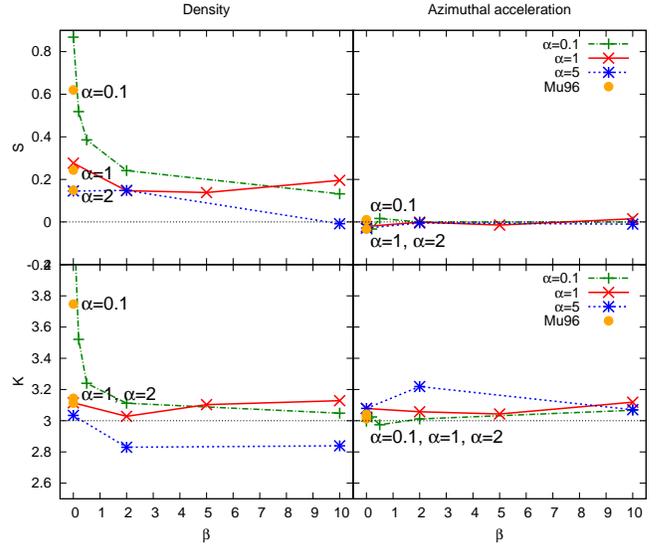}
\caption{\emph{Higher order moments}. 
Skewness (top) and kurtosis (bottom) 
 of density distribution (left) and azimuthal acceleration 
distribution (right).
}
\label{fig12b}
\end{figure}

\underline{Effect of artificial viscosity}.
For a given $\alpha$, larger values of $\beta$ result
in values of the skewness of the density  PDF
closer to the Gaussian value $S=0$ (top left plot in Fig.~\ref{fig12b}). 
The same effect is observed when $\beta$ is kept constant
and $\alpha$ increased.
All other quantities show Gaussian values independently of
the AV parameter used, an example is given by
the azimuthal component of the acceleration 
(top right plot in Fig.~\ref{fig12b}).
Deviations of the kurtosis from the Gaussian value 
$K=3$ are qualitatively 
the same as for the skewness. 

The simulations with the \citetalias{Murray1996} artificial viscosity 
show systematically lower values of $S$ and $K$ with respect to 
simulations with the \citetalias{Monaghan1983} artificial viscosity and with the 
same $\alpha$ and $\beta=0$. The reason is that for the 
same $\alpha$, \citetalias{Murray1996} discs are more viscous than \citetalias{Monaghan1983}
discs since artificial viscosity is applied to all particles
in contrast to the \citetalias{Monaghan1983} case.
This trend is in agreement with that observed in \citetalias{Monaghan1983} simulations
with changing AV parameters: more viscous discs 
present lower $S$ and $K$.

\underline{Effect of resolution}.
The values of S and K only show small fluctuations when
the resolution is increased (not shown), suggesting that
convergence for these global quantities is present even for
a number of particle as low as $N \approx 2 \cdot 10^5$.

In conclusion, no significant deviation from a Gaussian distribution 
 is observed: intermittency is not present.
This result agrees with 
the intermediate turbulent Reynolds number
$\textrm{Re}_{\mathrm{\scriptstyle{T}}}$
proper to the present simulations 
(see  Sect.~\ref{sec:alphaRS}).
In fact intermittency is usually observed 
in very high Reynolds number flows \citep[e.g.][]{Frisch1996}.

\section{Discussion and conclusion}
\label{sec:conclusions}
We have presented the characterisation of
the global flow and of the statistical properties 
of the fluctuations in the velocity and density field
of the gas in \emph{SPH disc models}, 
which are gaseous discs simulated by means of the SPH method.

An essential tool for most of our results is
the new method we have introduced for the determination
of the \emph{effective viscosity} of three dimensional axisymmetric discs.
It consists in fitting the analytic expression for the radial 
velocity derived from two dimensional models
\citepalias[see e.g.][]{Takeuchi2002},
depending both on $R$ and $z$,
to the vertical velocity profile  extracted from the simulation
at different radial positions in the disc.

We have focused on the effects of changing the number 
of particles $N$ and the values of the two AV parameters
 $\alpha$ and $\beta$. The relevant results are summarised
in the following points:
\begin{enumerate}
\item We have confirmed and quantified 
the numerical role of the number of particles $N$,
which contributes to the numerical component of the 
\emph{SPH fluctuations} (see its definition in Sect.~\ref{sec:intro}).
In fact, for all studied quantities, an 
increase in the resolution of the simulation 
(increase in the number of particles)
has the standard numerical role of convergence towards 
the physical solution, and also leads
to a more extended cascade in the power spectrum 
of velocity fluctuations
(smaller scales and therefore higher wavenumbers are resolved).
\item More significantly, we have found that the artificial viscosity
(through the two AV parameters $\alpha$ and $\beta$) 
contributes both to the numerical and to the physical component of 
the  \emph{SPH fluctuations}, in contrast to the behaviour of $N$. 
\end{enumerate}

There exists a relationship between artificial viscosity and the 
following three quantities:
(a)  physical fluctuations, quantified by
$\alpha_{\mathrm{\scriptstyle{RS}}}$, (b) the effective viscosity of the disc, quantified by
$\alpha_{\mathrm{\scriptstyle{2D}}}$ and (c) the numerical noise. 
In the hypothesis of an ideal noise-free numerical scheme,
we expect that at 
low $(\alpha,\beta)$ the effective viscosity is due to turbulent
viscosity, directly produced by physical fluctuations.
At intermediate $(\alpha,\beta)$  the effective viscosity 
has a contribution not only from physical fluctuations but also from 
the AV term,
which reproduces the effects of turbulence
without directly modeling eddies and vortices
in the spirit of the $\alpha_{\mathrm{\scriptstyle{SS}}}$-disc model.
Finally, in the high $(\alpha,\beta)$  range, physical fluctuations
are negligible and the  effective viscosity is dominated by
the AV term.
However, numerical schemes are affected by numerical noise,
which masks physical fluctuations. 

In our case,  noise is related to the AV parameters: 
only for $(\alpha,\beta)$ above a threshold is the noise
reduced enough to allow physical fluctuations to emerge.
We have identified the threshold at $(\alpha,\beta) \approx$ (1,2).
 In fact, in the discs simulated above this threshold, 
we have observed a sharp transition both
of the \emph{global} and of the \emph{local} behaviour.
\begin{itemize}
\item Concerning the global behaviour, we have found that with increasing artificial viscosity the global flow of the gaseous disc evolves from a chaotic 
radial velocity structure to a more ordered 
meridional circulation pattern characterised by larger accretion rates
and with the Reynolds number
$\textrm{Re}_{\mathrm{\scriptstyle{eff}}}$, associated to the estimated
effective viscosity $\alpha_{\mathrm{\scriptstyle{2D}}}$, 
below the turbulent limit for all
simulations (however we observe that some of the high resolution 
simulations are approaching it).
\item Concerning the local behaviour, 
we observe that, in parallel to the onset of 
meridional circulation, the average Reynolds stress
switches abruptly from positive to negative values and
converges to a value that 
contributes at the 10 per cent level
to the effective viscosity
($\alpha_{\mathrm{\scriptstyle{RS}}} \approx 0.1 \alpha_{\mathrm{\scriptstyle{2D}}}$),
with an associated turbulent Reynolds number
$\textrm{Re}_{\mathrm{\scriptstyle{T}}}$ around the turbulent limit.
In particular, we have found that with increasing artificial viscosity,
physical fluctuations have the following properties:
\begin{itemize}
\item velocity fluctuations  become more and more subsonic and shift
from radial to azimuthal anisotropy;
\item a cascade appears in the power spectrum of velocity fluctuations,
which tends towards a Kolmogorov-like spectrum, 
particularly for the radial component;
\item turbulent diffusion appears at the same time as the cascade
and then the diffusion coefficient tends to decrease.
Diffusion tends to be damped
by the onset of meridional circulation;
\item  the presence of meridional circulation leads to
 a shift of the peak of the azimuthal acceleration PDF;
\item  none of the studied SPH disc models presents fluctuations
characterised by intermittency.
\end{itemize}
\end{itemize}

With these results we can give a first answer to 
the question raised in the Introduction:
can SPH disc models correctly \emph{reproduce} both 
the observed properties of PPD and the expected effect of
turbulence?

In the case of \emph{SPH disc models} with the 
\citetalias{Monaghan1983} artificial viscosity 
we have found that for AV parameters 
above a threshold that approximately corresponds
to the standard values ($\alpha \approx 1$ and $\beta \approx 2$),
the effective viscosity
 mainly represents viscosity due to turbulence
 (of unknown origin) in the spirit of the \citetalias{Shakura1973}
$\alpha$-discs where intermediate and large scale 
turbulent eddies and vortices 
 are not resolved. However, at the smallest resolved scales,
 fluctuations with properties similar to those 
present in turbulent flows  have been identified.
A meridional circulation pattern
and turbulent-like fluctuations
coexist in SPH disc models.
The former shapes the global scale of the flow,
reproducing the observed \emph{mass accretion rate} onto the star
while the latter is present locally at the smallest resolved scales
 and contributes to only 10 per cent
of the effective viscosity of the disc.
The resolved physical fluctuations reproduce
 the main \emph{effects} expected in a turbulent 3D disc.
In fact, the simulated discs are characterised by:
a \emph{subsonic} fluctuating
velocity field presenting a turbulent-like \emph{diffusion} 
and a small Kolmogorov-like \emph{cascade} in the power spectrum
of velocity fluctuations at the smallest resolved scales
(see e.g. simulations S7 and S8).
These simulations  present values
of the Reynolds stress and of the diffusion coefficient that are very close 
to those observed in MHD accretion discs
(where the source of turbulence is the MRI).
For values of $\alpha$ and $\beta$ below the threshold,
the numerical noise dominates and masks physical fluctuations,
 these models should be avoided.

A direct detection of turbulent eddies would require
a reduction of the AV term in order to  
increase the contribution of  the Reynolds stress.
However, we have shown that for the two
AV implementations considered here 
the numerical noise dominates
and impedes the growth of eddies in the velocity field 
when the levels of AV are below the threshold:
in such cases the positive effects of the AV terms 
(particularly of the $\beta$ term) in
correctly describing SPH particles trajectories and 
in avoiding particle interpenetration are missing.
Therefore, in order to reach higher Reynolds numbers 
it is necessary both to reduce
the effective viscosity of the disc \emph{without} increasing the 
numerical noise
and to increase the resolution.
This is not possible with the two artificial viscosity 
implementation we are analysing in this work 
(since reducing the AV coefficients leads to an increase of the 
numerical noise).
Recently developed artificial viscosity switches 
\citep[e.g][]{Morris1997,Cullen2010}
could help in this direction. However, it is still not possible to
simulate fluids at the high Reynolds numbers (Re$_{\textrm{eff}}$~$\gg 10^3$)
expected in astrophysical flows with the current tools and computer power.

We should stress that simulating very high Reynolds number discs 
would be necessary if one wanted to investigate the
possibility of the existence of pure hydrodynamics turbulence in 3D
accretion discs. (We remind the reader that 1D accretion discs are
known to be stable with respect to hydrodynamics instabilities due to the
Rayleigh criterion, see e.g. \citealt{Armitage2007}.)
However, for our purpose of reproducing the effects of turbulence
produced by an unknown source, the effective viscosity
combined with a low and possibly intermediate Reynols stress is enough.
In fact, these models represent a starting point for the 
study of dust dynamics, which is affected by 
physical fluctuations at small scales and meridional
circulation at large scales.

In conclusion, in \emph{SPH disc models}, which a priori
include only the ingredients of \emph{star gravity} and 
\emph{physical-like-viscosity},
we have found that the artificial viscosity term,
in addition to modeling a physical-like bulk and shear viscosity,
can also play the role of an \emph{implicit turbulence model}.
In our SPH disc models, where we do not add any initial turbulent velocity field, the implicit turbulence model
 sustains and organises the random fluctuations
present in the initial conditions of the disc.
This result is in agreement with recent indications that a
turbulence model is implicitly present  in the SPH scheme
used to simulate homogeneous and isotropic turbulence 
with periodic  boundary conditions 
\citep[see e.g.][]{Shi2012,Monaghan2011,Ellero2010}.
Here we have analysed the more complex case of
systems without boundary conditions and 
where anisotropic turbulence is expected.

The effects of the density and sound speed profiles on the
global and statistical properties of the gas
as well as those of the initial set up,
of different SPH kernels
and of different artificial viscosity implementations
will be addressed in more detail in a future work.

\section*{Acknowledgments}
The authors wish to thank Roland Speith and Guillaume Laibe
for fruitful discussions.
This research was supported by the Agence Nationale de la
Recherche (ANR) of France through contract ANR-07-BLAN-0221
and was conducted within the Lyon Institute of Origins under
grant ANR-10-LABX-66.
Simulations presented in this work were run at the Service Commun
de Calcul Intensif (SCCI) de l'Observatoire de Grenoble, France.
Figure~\ref{fig1} was made using the {\sc splash} software
package \citep{Price2007}.

\bibliographystyle{mn2e}
\bibliography{bibliography}{}

\begin{thebibliography}{}

\bibitem[\protect\citeauthoryear{{Andrews}, {Wilner}, {Hughes}, {Qi} \&
  {Dullemond}}{{Andrews} et~al.}{2009}]{Andrews2009}
{Andrews} S.~M.,  {Wilner} D.~J.,  {Hughes} A.~M.,  {Qi} C.,    {Dullemond}
  C.~P.,  2009, \apj, 700, 1502

\bibitem[\protect\citeauthoryear{{Armitage}}{{Armitage}}{2007}]{Armitage2007}
{Armitage} P.~J.,  2007, arXiv:0701485

\bibitem[\protect\citeauthoryear{{Balbus} \& {Hawley}}{{Balbus} \&
  {Hawley}}{1991}]{Balbus1991}
{Balbus} S.~A.,  {Hawley} J.~F.,  1991, \apj, 376, 214

\bibitem[\protect\citeauthoryear{{Barri{\`e}re-Fouchet}, {Gonzalez}, {Murray},
  {Humble} \& {Maddison}}{{Barri{\`e}re-Fouchet}
  et~al.}{2005}]{Barri`ere-Fouchet2005}
{Barri{\`e}re-Fouchet} L.,  {Gonzalez} J.-F.,  {Murray} J.~R.,  {Humble} R.~J.,
     {Maddison} S.~T.,  2005, \aap, 443, 185

\bibitem[\protect\citeauthoryear{{Bertin}}{{Bertin}}{2000}]{Bertin2000}
{Bertin} G.,  2000, {Dynamics of Galaxies}

\bibitem[\protect\citeauthoryear{{Carballido}, {Stone} \&
  {Turner}}{{Carballido} et~al.}{2008}]{Carballido2008}
{Carballido} A.,  {Stone} J.~M.,    {Turner} N.~J.,  2008, \mnras, 386, 145

\bibitem[\protect\citeauthoryear{{Cullen} \& {Dehnen}}{{Cullen} \&
  {Dehnen}}{2010}]{Cullen2010}
{Cullen} L.,  {Dehnen} W.,  2010, \mnras, 408, 669

\bibitem[\protect\citeauthoryear{{Cuzzi}, {Dobrovolskis} \& {Champney}}{{Cuzzi}
  et~al.}{1993}]{Cuzzi1993}
{Cuzzi} J.~N.,  {Dobrovolskis} A.~R.,    {Champney} J.~M.,  1993, \icarus, 106,
  102

\bibitem[\protect\citeauthoryear{{Cuzzi}, {Hogan} \& {Shariff}}{{Cuzzi}
  et~al.}{2008}]{Cuzzi2008}
{Cuzzi} J.~N.,  {Hogan} R.~C.,    {Shariff} K.,  2008, \apj, 687, 1432

\bibitem[\protect\citeauthoryear{Ellero, Espa\~nol \& Adams}{Ellero
  et~al.}{2010}]{Ellero2010}
Ellero M.,  Espa\~nol P.,    Adams N.~A.,  2010, Phys. Rev. E, 82, 046702

\bibitem[\protect\citeauthoryear{{Flock}, {Dzyurkevich}, {Klahr}, {Turner} \&
  {Henning}}{{Flock} et~al.}{2012}]{Flock2011a}
{Flock} M.,  {Dzyurkevich} N.,  {Klahr} H.,  {Turner} N.,    {Henning} T.,
  2012, \apj, 744, 144

\bibitem[\protect\citeauthoryear{{Flock}, {Dzyurkevich}, {Klahr}, {Turner} \&
  {Henning}}{{Flock} et~al.}{2011}]{Flock2011}
{Flock} M.,  {Dzyurkevich} N.,  {Klahr} H.,  {Turner} N.~J.,    {Henning} T.,
  2011, \apj, 735, 122

\bibitem[\protect\citeauthoryear{{Frisch}}{{Frisch}}{1996}]{Frisch1996}
{Frisch} U.,  1996, {Turbulence}

\bibitem[\protect\citeauthoryear{{Fromang}, {Lyra} \& {Masset}}{{Fromang}
  et~al.}{2011}]{Fromang2011}
{Fromang} S.,  {Lyra} W.,    {Masset} F.,  2011, \aap, 534, A107

\bibitem[\protect\citeauthoryear{{Fromang} \& {Nelson}}{{Fromang} \&
  {Nelson}}{2006}]{Fromang2006}
{Fromang} S.,  {Nelson} R.~P.,  2006, \aap, 457, 343

\bibitem[\protect\citeauthoryear{{Fromang} \& {Nelson}}{{Fromang} \&
  {Nelson}}{2009}]{Fromang2009}
{Fromang} S.,  {Nelson} R.~P.,  2009, \aap, 496, 597

\bibitem[\protect\citeauthoryear{{Fromang} \& {Papaloizou}}{{Fromang} \&
  {Papaloizou}}{2006}]{Fromang2006a}
{Fromang} S.,  {Papaloizou} J.,  2006, \aap, 452, 751

\bibitem[\protect\citeauthoryear{{Guilloteau}, {Dutrey}, {Wakelam}, {Hersant},
  {Semenov}, {Chapillon}, {Henning} \& {Pi{\'e}tu}}{{Guilloteau}
  et~al.}{2012}]{Guilloteau2012}
{Guilloteau} S.,  {Dutrey} A.,  {Wakelam} V.,  {Hersant} F.,  {Semenov} D.,
  {Chapillon} E.,  {Henning} T.,    {Pi{\'e}tu} V.,  2012, \aap, 548, A70

\bibitem[\protect\citeauthoryear{{Hartmann}, {Calvet}, {Gullbring} \&
  {D'Alessio}}{{Hartmann} et~al.}{1998}]{Hartmann1998}
{Hartmann} L.,  {Calvet} N.,  {Gullbring} E.,    {D'Alessio} P.,  1998, \apj,
  495, 385

\bibitem[\protect\citeauthoryear{{Hughes}, {Wilner}, {Andrews}, {Qi} \&
  {Hogerheijde}}{{Hughes} et~al.}{2011}]{Hughes2011}
{Hughes} A.~M.,  {Wilner} D.~J.,  {Andrews} S.~M.,  {Qi} C.,    {Hogerheijde}
  M.~R.,  2011, \apj, 727, 85

\bibitem[\protect\citeauthoryear{{Jacquet}}{{Jacquet}}{2013}]{Jacquet2013}
{Jacquet} E.,  2013, \aap, in press (arXiv:1301.5817)

\bibitem[\protect\citeauthoryear{{King}, {Pringle} \& {Livio}}{{King}
  et~al.}{2007}]{King2007}
{King} A.~R.,  {Pringle} J.~E.,    {Livio} M.,  2007, \mnras, 376, 1740

\bibitem[\protect\citeauthoryear{{Klahr} \& {Bodenheimer}}{{Klahr} \&
  {Bodenheimer}}{2003}]{Klahr2003}
{Klahr} H.~H.,  {Bodenheimer} P.,  2003, \apj, 582, 869

\bibitem[\protect\citeauthoryear{{Laibe}, {Gonzalez} \& {Maddison}}{{Laibe}
  et~al.}{2012}]{Laibe2012}
{Laibe} G.,  {Gonzalez} J.-F.,    {Maddison} S.~T.,  2012, \aap, 537, A61

\bibitem[\protect\citeauthoryear{{Lattanzio}, {Monaghan}, {Pongracic} \&
  {Schwarz}}{{Lattanzio} et~al.}{1985}]{Lattanzio1985}
{Lattanzio} J.~C.,  {Monaghan} J.~J.,  {Pongracic} H.,    {Schwarz} M.~P.,
  1985, \mnras, 215, 125

\bibitem[\protect\citeauthoryear{Lodato}{Lodato}{2008}]{Lodato2008}
Lodato G.,  2008, New Astronomy Reviews, 52, 21

\bibitem[\protect\citeauthoryear{{Lodato} \& {Price}}{{Lodato} \&
  {Price}}{2010}]{Lodato2010}
{Lodato} G.,  {Price} D.~J.,  2010, \mnras, 405, 1212

\bibitem[\protect\citeauthoryear{{Lovelace}, {Li}, {Colgate} \&
  {Nelson}}{{Lovelace} et~al.}{1999}]{Lovelace1999}
{Lovelace} R.~V.~E.,  {Li} H.,  {Colgate} S.~A.,    {Nelson} A.~F.,  1999,
  \apj, 513, 805

\bibitem[\protect\citeauthoryear{{Lynden-Bell} \& {Pringle}}{{Lynden-Bell} \&
  {Pringle}}{1974}]{Lynden-Bell1974}
{Lynden-Bell} D.,  {Pringle} J.~E.,  1974, \mnras, 168, 603

\bibitem[\protect\citeauthoryear{{Meglicki}, {Wickramasinghe} \&
  {Bicknell}}{{Meglicki} et~al.}{1993}]{Meglicki1993}
{Meglicki} Z.,  {Wickramasinghe} D.,    {Bicknell} G.~V.,  1993, \mnras, 264,
  691

\bibitem[\protect\citeauthoryear{{Meru} \& {Bate}}{{Meru} \&
  {Bate}}{2012}]{Meru2012}
{Meru} F.,  {Bate} M.~R.,  2012, \mnras, 427, 2022

\bibitem[\protect\citeauthoryear{{Monaghan}}{{Monaghan}}{1992}]{Monaghan1992}
{Monaghan} J.~J.,  1992, \araa, 30, 543

\bibitem[\protect\citeauthoryear{{Monaghan}}{{Monaghan}}{2005}]{Monaghan2005}
{Monaghan} J.~J.,  2005, Reports on Progress in Physics, 68, 1703

\bibitem[\protect\citeauthoryear{Monaghan}{Monaghan}{2011}]{Monaghan2011}
Monaghan J.~J.,  2011, European Journal of Mechanics - B/Fluids, 30, 360

\bibitem[\protect\citeauthoryear{Monaghan \& Gingold}{Monaghan \&
  Gingold}{1983}]{Monaghan1983}
Monaghan J.~J.,  Gingold R.~A.,  1983, Journal of Computational Physics, 52,
  374

\bibitem[\protect\citeauthoryear{{Monaghan} \& {Lattanzio}}{{Monaghan} \&
  {Lattanzio}}{1985}]{Monaghan1985}
{Monaghan} J.~J.,  {Lattanzio} J.~C.,  1985, \aap, 149, 135

\bibitem[\protect\citeauthoryear{Morris \& Monaghan}{Morris \&
  Monaghan}{1997}]{Morris1997}
Morris J.~P.,  Monaghan J.~J.,  1997, Journal of Computational Physics, 136, 41

\bibitem[\protect\citeauthoryear{{Murray}}{{Murray}}{1996}]{Murray1996}
{Murray} J.~R.,  1996, \mnras, 279, 402

\bibitem[\protect\citeauthoryear{{Nelson}, {Gressel} \& {Umurhan}}{{Nelson}
  et~al.}{2012}]{Nelson2012}
{Nelson} R.~P.,  {Gressel} O.,    {Umurhan} O.~M.,  2012, ArXiv e-prints

\bibitem[\protect\citeauthoryear{Pope}{Pope}{2000}]{Pope2000}
Pope S.~B.,  2000, {Turbulent Flows}.
Cambridge University Press

\bibitem[\protect\citeauthoryear{{Price}}{{Price}}{2007}]{Price2007}
{Price} D.~J.,  2007, \pasa, 24, 159

\bibitem[\protect\citeauthoryear{{Price}}{{Price}}{2012}]{Price2012}
{Price} D.~J.,  2012, \mnras, 420, L33

\bibitem[\protect\citeauthoryear{{Price} \& {Federrath}}{{Price} \&
  {Federrath}}{2010}]{Price2010}
{Price} D.~J.,  {Federrath} C.,  2010, \mnras, 406, 1659

\bibitem[\protect\citeauthoryear{{Pringle}}{{Pringle}}{1981}]{Pringle1981}
{Pringle} J.~E.,  1981, Annual review of astronomy and astrophysics, 19, 137

\bibitem[\protect\citeauthoryear{{Rice}, {Armitage}, {Mamatsashvili}, {Lodato}
  \& {Clarke}}{{Rice} et~al.}{2011}]{Rice2011}
{Rice} W.~K.~M.,  {Armitage} P.~J.,  {Mamatsashvili} G.~R.,  {Lodato} G.,
  {Clarke} C.~J.,  2011, \mnras, pp 1535--+

\bibitem[\protect\citeauthoryear{{Rice}, {Lodato} \& {Armitage}}{{Rice}
  et~al.}{2005}]{Rice2005}
{Rice} W.~K.~M.,  {Lodato} G.,    {Armitage} P.~J.,  2005, \mnras, 364, L56

\bibitem[\protect\citeauthoryear{{Sargent}, {Forrest}, {Tayrien}, {McClure},
  {Watson}, {Sloan}, {Li}, {Manoj}, {Bohac}, {Furlan}, {Kim} \&
  {Green}}{{Sargent} et~al.}{2009}]{Sargent2009}
{Sargent} B.~A.,  {Forrest} W.~J.,  {Tayrien} C.,  {McClure} M.~K.,  {Watson}
  D.~M.,  {Sloan} G.~C.,  {Li} A.,  {Manoj} P.,  {Bohac} C.~J.,  {Furlan} E.,
  {Kim} K.~H.,    {Green} J.~D.,  2009, \apjs, 182, 477

\bibitem[\protect\citeauthoryear{{Sch{\"a}fer}}{{Sch{\"a}fer}}{2005}]{Schaefer2005}
{Sch{\"a}fer} C.,  2005, PhD thesis, Eberhard-Karls Universit{\"a}t
  T{\"u}bingen

\bibitem[\protect\citeauthoryear{{Shakura} \& {Sunyaev}}{{Shakura} \&
  {Sunyaev}}{1973}]{Shakura1973}
{Shakura} N.~I.,  {Sunyaev} R.~A.,  1973, in {H.~Bradt \& R.~Giacconi} ed., X-
  and Gamma-Ray Astronomy Vol.~55 of IAU Symposium, {Black Holes in Binary
  Systems: Observational Appearances}.
pp 155--+

\bibitem[\protect\citeauthoryear{Shi, Ellero \& Adams}{Shi
  et~al.}{2012}]{Shi2012}
Shi Y.,  Ellero M.,    Adams N.~A.,  2012, Phys. Rev. E, 85, 036708

\bibitem[\protect\citeauthoryear{{Takeuchi} \& {Lin}}{{Takeuchi} \&
  {Lin}}{2002}]{Takeuchi2002}
{Takeuchi} T.,  {Lin} D.~N.~C.~.,  2002, \apj, 581, 1344

\bibitem[\protect\citeauthoryear{Violeau \& Issa}{Violeau \&
  Issa}{2007}]{VioleauIssa2007}
Violeau D.,  Issa R.,  2007, International Journal for Numerical Methods in
  Fluids, 53, 277

\end{thebibliography}

\appendix

\section[]{Diagnostic details}
\subsection[]{The turbulent viscosity coefficient $\nu_{\mathrm{\scriptstyle{T}}}$}
\label{app:turbulentVisc}
From the turbulent viscosity hypothesis \citep{Pope2000},
the Reynolds stresses are written in Cartesian coordinates as follows:
\begin{equation}
\langle u_i u_j \rangle = - 2 \nu_{\mathrm{\scriptstyle{T}}} \dot{\varepsilon}_{ij} - \frac{2}{3} k \delta_{ij},
\end{equation}
where $\dot{\varepsilon}_{ij}$ is the rate of strain:
\begin{equation}
\dot{\varepsilon}_{ij} = \frac{1}{2} \left( \frac{\partial \langle v_i \rangle}{\partial x_j} + \frac{\partial \langle v_j \rangle}{\partial x_i} \right) 
\end{equation}
and $k$ the turbulent kinetic energy:
\begin{equation}
k = \frac{1}{2} \langle u_i u_i \rangle.
\end{equation}
In classical accretion disc theory,
 only shear viscosity is relevant.
Therefore the only non-vanishing component
of the stress tensor is the $R \theta$ component \citep[see e.g][]{Lodato2008},
which in cylindrical coordinates becomes:
\begin{equation}
\label{eq:stress}
\langle u_R u_{\theta} \rangle = - \nu_{\mathrm{\scriptstyle{T}}} \left[ R \frac{\partial \left( \langle v_{\theta} \rangle / R \right)}{\partial R} + \frac{1}{R} \frac{\partial \langle v_R \rangle}{\partial \theta} \right].
\end{equation}
Given the axisymmetry of the disc, the second term on the right hand side 
of Eq.~\ref{eq:stress} vanishes and the turbulent viscosity
coefficient becomes:
\begin{equation}
\label{eq:viscosity}
  \nu_{\mathrm{\scriptstyle{T}}} = - \frac{\langle u_R u_{\theta} \rangle}{R \left( \langle v_{\theta} \rangle / R \right)'},
\end{equation}
with $(\cdot)' \equiv \partial / \partial R$.

\subsection[]{The $\alpha_{\mathrm{\scriptstyle{RS}}}$ coefficient} 
\label{app:alphaRS}
The \citetalias{Shakura1973} coefficient corresponding to the viscosity of Sect.~\ref{app:turbulentVisc}
is defined by the relation $\nu_{\mathrm{\scriptstyle{T}}} = \alpha_{\mathrm{\scriptstyle{RS}}} c_\mathrm{s} H$ that combined with Eq.~\ref{eq:viscosity} gives:
\begin{equation}
\alpha_{\mathrm{\scriptstyle{RS}}} = - \frac{\langle u_R u_{\theta} \rangle}{c_\mathrm{s} H R \left( \langle v_{\theta} \rangle / R \right)'}
\end{equation}

In the case of \emph{quasi-Keplerian} discs we have:  $\langle v_{\theta} \rangle$~$\approx$~$v_\mathrm{k}$~=~$R \Omega_\mathrm{k}$, with $v_\mathrm{k}$ the Keplerian velocity and $\Omega_\mathrm{k}=\sqrt{GM} r^{-3/2}$ the corresponding  angular velocity.
For \emph{thin} discs, the approximation $r \approx R$ holds, with $r$ the radial spherical component and   $R$ the radial cylindrical component.

Remembering that the scale height of the disc is related to the sound speed and to the Keplerian angular velocity by $H=c_\mathrm{s}/\Omega_\mathrm{k}$, the $\alpha_{\mathrm{\scriptstyle{RS}}}$ coefficient takes the form:
\begin{equation}
\alpha_{\mathrm{\scriptstyle{RS}}} = - \frac{\langle u_R u_{\theta} \rangle}{R (\Omega_\mathrm{k})'} \frac{\Omega_\mathrm{k}}{c_\mathrm{s}^2} = \frac{2}{3} \frac{\langle u_R u_{\theta} \rangle}{c_\mathrm{s}^2}.
\end{equation}

\subsection[]{The power spectrum, PDFs, $S$ and $K$}
\label{app:PDFsEtAl}
We consider a ring made of $N_g$ points and centred at the origin of the disc, with selected radius $R_\mathrm{s}$ and height $z_s$.
The density and the smoothing length of each grid point are computed by means of an iterating procedure using the two coupled equations proper to the SPH scheme: 
\begin{equation}
\rho_a = \sum_b m_b W_{ab}(r_{ab},h_a); \qquad h_a = \eta \left( \frac{m_a}{\rho_a}  \right)^{1/3},
\end{equation}
where the subscript $a$ refers to the grid point and the subscript $b$ to its neighbours.

Once the smoothing length is known, the value of the desired quantities 
(e.g. velocity components for the power spectrum, density and azimuthal acceleration for the PDFs and higher order moments) at each grid point are computed by the SPH smoothing technique.
Now the one dimensional list of values is known and used to compute the power spectrum of velocity components, the PDFs and the corresponding $S$ and $K$ coefficients.

This procedure is applied to each simulation snapshot, then values are
averaged in time (for 15.9 orbits at 100 au).
For $S$ and $K$ the standard deviation of the distribution of values in time has also been considered.

\section[]{Procedure for fitting the effective viscosity}
\label{app:fit}
The effective viscosity $\alpha_{\scriptstyle \textrm{2D}}$ in the disc  is derived by fitting Eq.~\ref{eq:vr} to the data computed for a given snapshot of the simulation by means of the following three steps:
\begin{enumerate}
\item Computation of data from the selected simulation snapshot: 
radial profile  $\Sigma(R)$ of the surface density, vertical profiles $\rho(z)$ and $v_R(z)$ of the volumetric density and radial velocity at radial position $R$.
\item Determination of the parameters present in  Eq.~\ref{eq:vr}. The surface density power law $p$ at the selected location $R$ is derived by fit of the surface density profile. The scale height $H$ of the disc at $R$ and $R_0$ is derived by fit of the relative vertical density profiles. Note that $q$ is constant, since simulations are locally isothermal, $R_0$ is the length unit and $c_{s_0}=H_0=H(R_0)$ since $R_0=G=M=1$.
\item Determination of $\alpha_{\scriptstyle \textrm{2D}}$ by fit to the $v_R$ data. 
\end{enumerate}

All fits are performed by the general least square method, with $\chi^2$ function defined by:
\begin{equation}
\chi^2=\sum_{i=1}^{N} \left[ \frac{y_i - f(x_i,a)}{\sigma_i}  \right]^2,
\end{equation}
where $(x_i,y_i)$ are the $N$ data points from simulations, $\sigma_i$ the associated errors and $f(x_i,a)$ the function to be fitted, which depends on the parameter $a$.

\label{lastpage}
\end{document}